\documentclass[aps,pre,superscriptaddress,twocolumn,nobibnotes]{revtex4-1}

\usepackage{graphicx}
\usepackage{todonotes}
\usepackage{amsmath}
\usepackage{amsmath}
\usepackage{upgreek}
\usepackage{color}
\usepackage{hyperref}

\usepackage{gensymb}

\usepackage{hyperref}

\renewcommand{\phi}{ \varphi }

\newcommand{\degC}[1]{{$^{\rm\circ}$}}

\newcommand{\revision}[1]{\textcolor{black}{#1}}
\begin{document}

\title{Self-Assembly of Liquid Crystals in Nanoporous Solids\\ for Adaptive Photonic Metamaterials}

\author{Kathrin Sentker}
\affiliation{Institut f\"ur Materialphysik und -technologie, 
    Technische Universit\"at Hamburg, Ei{\ss}endorferstr. 42, D-21073 Hamburg, Germany}
\author{Arda Yildirim}
\affiliation{Bundesanstalt f\"ur Materialforschung und -pr\"ufung, Unter den Eichen 87, D-12205 Berlin, Germany}
\author{Milena Lippmann}
\affiliation{Deutsches Elektronen Synchrotron, 
	Notkestra{\ss}e 85, D-22607 Hamburg, Germany}


\author{Arne W. Zantop}
\affiliation{Max-Planck-Institut f\"ur Dynamik und Selbstorganisation, 
    Am Fa{\ss}berg 17, D-37077 G\"ottingen, Germany}

\author{Florian Bertram}
\affiliation{Deutsches Elektronen Synchrotron, 
	Notkestra{\ss}e 85, D-22607 Hamburg, Germany}

\author{Tommy Hofmann}
\affiliation{Helmholtz-Zentrum Berlin f\"ur Materialien und Energie,
Hahn-Meitner-Platz 1, D-14109 Berlin, Germany}

\author{Oliver H. Seeck}
\affiliation{Deutsches Elektronen Synchrotron, 
	Notkestra{\ss}e 85, D-22607 Hamburg, Germany}

\author{Andriy V. Kityk}
\affiliation{Faculty of Electrical Engineering, Czestochowa University of Technology, Al.~Armii Krajowej 17, P-42-200 Czestochowa, Poland}

\author{Marco G. Mazza}
\affiliation{Max-Planck-Institut f\"ur Dynamik und Selbstorganisation, 
    Am Fa{\ss}berg 17, D-37077 G\"ottingen, Germany}
\affiliation{Interdisciplinary Centre for Mathematical Modelling and Department of Mathematical Sciences, Loughborough University, Loughborough, Leicestershire LE11 3TU, UK}
\author{Andreas Sch\"onhals}
\affiliation{Bundesanstalt f\"ur Materialforschung und -pr\"ufung, Unter den Eichen 87, D-12205 Berlin, Germany}

\author{Patrick Huber}
\email[]{patrick.huber@tuhh.de}
\affiliation{Institut f\"ur Materialphysik und -technologie, 
    Technische Universit\"at Hamburg, Ei{\ss}endorferstr. 42, D-21073 Hamburg, Germany}


\begin{abstract}{Nanoporous media exhibit structures significantly smaller than the wavelengths of visible light and can thus act as photonic metamaterials. Their optical functionality is not determined by the properties of the base materials, but rather by tailored, multiscale structures, in terms of precise pore shape, geometry, and orientation. Embedding liquid crystals in pore space provides additional opportunities to control light-matter interactions at the single-pore, meta-atomic scale. Here, we present temperature-dependent 3D reciprocal space mapping using synchrotron-based X-ray diffraction in combination with high-resolution birefringence experiments on disk-like mesogens (HAT6) imbibed in self-ordered arrays of parallel cylindrical pores 17 to 160\,nm across in monolithic anodic aluminium oxide (AAO). In agreement with Monte Carlo computer simulations we observe a remarkably rich self-assembly behaviour, unknown from the bulk state. It encompasses transitions between the isotropic liquid state and discotic stacking in linear columns as well as circular concentric ring formation perpendicular and parallel to the pore axis. These textural transitions underpin an optical birefringence functionality, tuneable in magnitude and in sign from positive to negative via pore size, pore surface-grafting and temperature. Our study demonstrates that the advent of large-scale, self-organised nanoporosity in monolithic solids along with confinement-controllable phase behaviour of liquid-crystalline matter at the single-pore scale provides a reliable and accessible tool to design materials with adjustable optical anisotropy, and thus offers versatile pathways to fine-tune polarisation-dependent light propagation speeds in materials. Such a tailorability is at the core of the emerging field of transformative optics, allowing, e.g., adjustable light absorbers and extremely thin metalenses.}
\end{abstract}

\date{25.11.2019}

\maketitle
\newpage
\section{Introduction}
When electromagnetic waves traverse a birefringent material their propagation speeds depend sensitively on their polarisation and propagation direction with respect to characteristic directions of the optical anisotropic medium. Birefringence and thus direction-dependent refractive indices are often associated with liquid crystals or anisotropic solid crystal materials. Amorphous materials are usually optically isotropic. However, if structural elements of nanoscale size, such as rods or pores with anisotropic shapes having uniform refractive index, are suspended in an optically isotropic material they are characterised by a so-called ``geometrical'' (or ``form'') birefringence. \revision{When the lattice spacing of these ``meta-atoms'' is much smaller than the wavelength of visible light the whole structure is described as a metamaterial \cite{Li2018, Kadic2019, Shaltout2019}.} A monolithic membrane traversed by a parallel array of cylindrical nanochannels represents such a metamaterial, see illustration in Fig.\,\ref{fig:BirefringenceIntro}. The advent of tailorable porosity at the nanoscale and thus of an adjustable ``meta-atomic'' structure offers new opportunities to tailor the optical anisotropy of such metamaterials. \revision{ Effective optical properties, not achievable by base materials, are possible \cite{Zheludev2012,Jalas2017, Lee2018, Nemati2018,ZhangM2018,Kadic2019, Shaltout2019}, specifically adjustable birefringence.} This is of particular relevance for important polarisation management components, such as waveplates and compensators, that play an essential part in modern optical communication systems \cite{Hu2019}. 
\iftrue
\begin{figure}[htbp]
 \centering
	{\includegraphics[width=0.9\columnwidth]{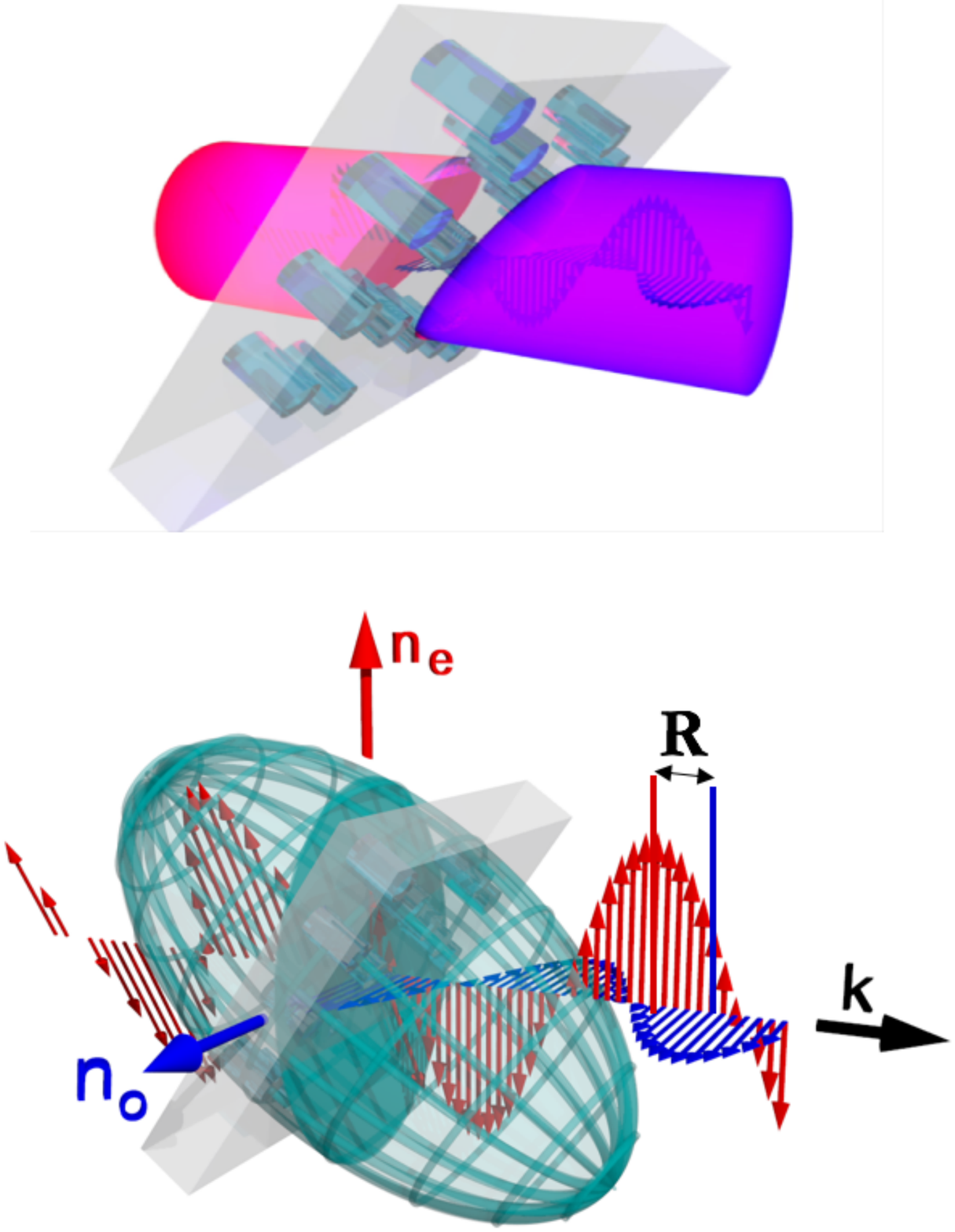}}
	\caption{\textbf{Optical birefringence of a nanoporous membrane.} (Top) A monolithic membrane traversed by an array of cylindrical nanopores in a linearly polarised laser beam. (Bottom) The incident electromagnetic wave is split up into two waves, i.e. the extraordinary (red) and ordinary wave (blue). These are polarised perpendicularly to each other. Because of the optical anisotropy, as indicated by the ellipsoid of refractive indices (indicatrix), they have different propagation speeds and thus encounter distinct refractive indices, the ordinary and extraordinary refractive index, $n_{\rm e}$ and $n_{\rm o}$, respectively. Here the extraordinary beam is slower than the ordinary beam resulting in a positive phase retardation $R$ of the electromagnetic waves after passing the pore array. This corresponds to a prolate indicatrix and a ''positive" optical birefringence, i.e. $\Delta n=n_{\rm e}-n_{\rm o}>0$. Note that for clarity only the electrical fields of the electromagnetic waves are shown.}
		\label{fig:BirefringenceIntro}
\end{figure}
\fi

A versatile route to fine-tune optics of porous materials is filling the pore space with a liquid crystal (LC) \cite{Matthias2005, Spengler2018}. In the bulk case LCs play a dominant role for the manipulation of optical anisotropy \cite{Buchnev2015}. By embedding LCs in nanoporous media the self-assembly and self-healing mechanisms can be used to design a new soft-hard metamaterial. The LC adds optical functionality, whereas the solid nanoporous material can guide the self-assembly and provides a mechanical stable, monolithic 3D scaffold structure \cite{Huber2015}. However, because of spatial and topological constraints, the phase and self-assembly behaviour of confined soft matter and in particular confined LCs may substantially deviate from the bulk state \cite{Ocko1986, Pershan1988, Crawford1996, Kutnjak2003,Binder2008,  Mazza2010, Chahine2010,Mueter2010, Araki2011,Cetinkaya2013, Calus2015, Schlotthauer2015, Ryu2016, Busch2017, Tran2017, Brumby2017, Zhang2019}.

Here, we employ archetypical disk-like mesogens (HAT6) with aromatic cores and aliphatic side chains as functional fillings of nanoporous anodic aluminium oxide (AAO) membranes, see Fig.~\ref{fig:HAT6AAOSEM}. The disk-like mesogens stack up in potentially ''infinitely long'' columns with short-range positional order, which arrange in a long-range two-dimensional lattice leading to discotic columnar liquid crystals (DLCs). Because of overlapping $\pi$-electrons of the aromatic cores DLCs exhibit a high one-dimensional charge carrier mobility along the columnar axes \cite{Chandrasekhar1990, Oswald2005, Feng2009, Sergeyev2007,Bisoyi2010,Kumar2010, Woehrle2015, Bisoyi2019}. Therefore, they are of high interest as 1-D semiconductors with possible applications in sensing, light harvesting or emission, and for molecular electronic components \cite{Woehrle2015}. Depending on their collective orientational and translational order they should also significantly alter the effective optical properties of nanoporous media.
	
Previous experimental studies of DLCs confined in cylindrical nanopores have revealed a remarkably complex, sometimes very surprising self-assembly behaviour \cite{Schmidt-Mende2001,Steinhart2005,Duran2012,Kityk2014, Calus2015, Ryu2016, Zhang2017,  Bisoyi2019}. Experimental studies on HAT6 in AAO indicate that for planar anchoring, i.e., the molecules lie flat on the hydrophilic pore walls, no long-range intercolumnar hexagonal order is detectable for pore diameters $d$ below 20\,nm \cite{Cerclier2012}. The columns stack in radial direction, see Fig.\,\ref{fig:MCIllustHydroPhilicPhobic}(a) for the illustration of such a configuration found in Monte Carlo (MC) simulations \cite{Zantop2015}. Additionally, Zhang et al. have found that the columns adopt a ``logpile'' configuration with parallel columns crossing the pores perpendicular to their axis for larger pores, see Fig.\,\ref{fig:MCIllustHydroPhilicPhobic}(b). Despite expectations, an axial arrangement of columns for edge-on anchoring was not found for neat HAT6 even in the smallest pore diameter of 20\,nm studied so far \cite{Zhang2015, Zhang2017}. The columns form rather circular concentric (CC) bent arrangements as depicted in the MC simulation snapshot, see Fig.\,\ref{fig:MCIllustHydroPhilicPhobic}(c) \cite{Zhang2014, Zhang2015, Zantop2015, Sentker2018}. An axial arrangement, see Fig.\,\ref{fig:MCIllustHydroPhilicPhobic}(d), which is of interest with respect to electronic applications, could only be achieved in DLCs with increased core-rigidity compared to HAT6 \cite{Zhang2012} or by non-circular channel cross-sections \cite{Zhang2019}.

In the following, temperature-dependent 2D X-ray diffraction and optical polarimetry reveal the evolution of collective orientational and translational order of HAT6 in self-organised hexagonally arranged cylindrical hydrophilic and hydrophobic AAO pores, see Fig.\ref{fig:HAT6AAOSEM}. Moreover, the resulting effective optical anisotropy, \textit{i.e.} birefringence, and thus the effective optics of the liquid-crystal-infused array of cylindrical nanochannels is analysed and  related to the complex liquid-crystalline texture formation within the pores.

\section{Experimental}
\subsection{Sample Preparation}
HAT6, see chemical structure in Fig.\,\ref{fig:HAT6AAOSEM}, is filled into AAO membranes via spontaneous imbibition \cite{Gruener2011}. The AAO membranes were purchased from Smart Membranes GmbH (SM), InRedox LLC (IR) as well as provided by the research group of M. Steinhart at University of Osnabr\"uck (OS). The porosities $P$ and pore diameters $d$ are checked by N$_2$-sorption isotherms at $T=$\,77\,K and vary from $P=$\,7.6\,\% to $P=$\,36.5\,\% and $d =$\,17.1\,nm to $d =$\,161\,nm, see Supplementary Tab.\,1. Naturally the AAO surface enforces face-on anchoring of DLCs. Chemically modification with octadecylphosphonic acid (ODPA) as described in Ref.\,\cite{Gri2011} renders the surface hydrophobic and thus enforces edge-on anchoring \cite{Yildirim2019}. \revision{The length of the ODPA molecule is $\approx$ 2.2\,nm and can thus result in a corresponding pore radius reduction. However, the grafting-layer density is not known in AAO and the filling with the liquid-crystalline melt could push the ODPA chains to the pore wall. Therefore, we state also for the surface-grafted AAO membranes the pore size and porosity as determined for the native membranes.} After outgassing at 200$^{\rm\circ}$C the membranes are filled with HAT6 by spontaneous imbibition for 48\,h at a temperature slightly above the bulk isotropic-nematic transition temperature $T^{\rm\rm\text{ci}}_{\rm\text{bulk}} = 371$\,K inside a glove box. Afterwards the left over material on top of the membrane is carefully removed with a razor blade. Prior and after filling the membranes were weighted to determine the mass of the confined HAT6.


\iftrue
\begin{figure}[htbp]
	\centering
	\includegraphics[width=0.4\textwidth]{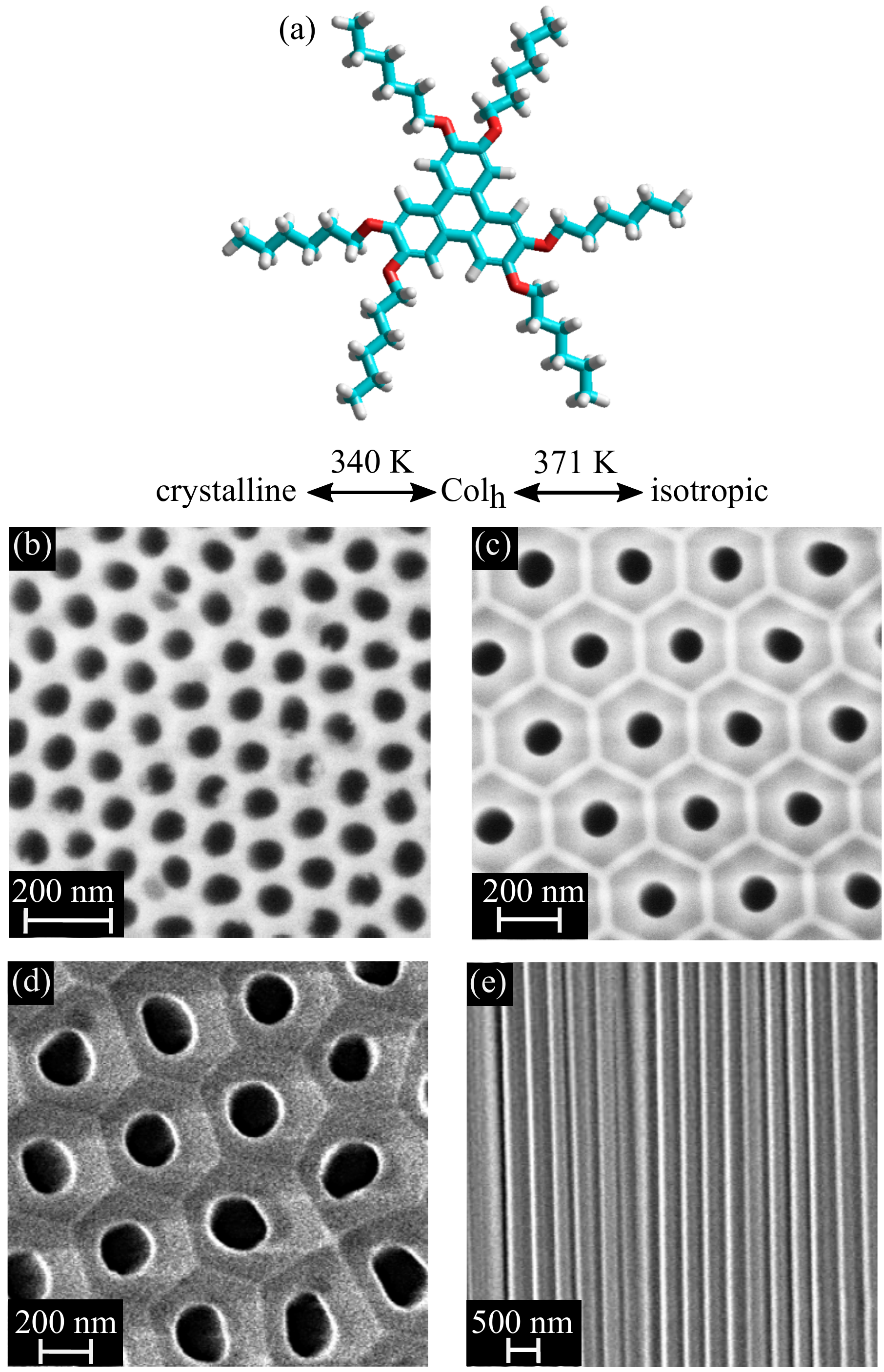}	
	\caption{\textbf{Structure of discotic liquid crystal and nanoporous solids.} (a) Chemical structure of discotic liquid crystal 2,3,6,7,10,11-hexakis(hexyloxy)triphenylene (HAT6) and phase transition temperatures in the bulk state. (b-d) Top and (e) side view electron micrographs of porous anodised aluminium oxide membranes with varying pore size.}
	\label{fig:HAT6AAOSEM}
\end{figure}
\fi

\iftrue
\begin{figure*}[htbp]
  \centering
		\includegraphics[width=1.8\columnwidth]{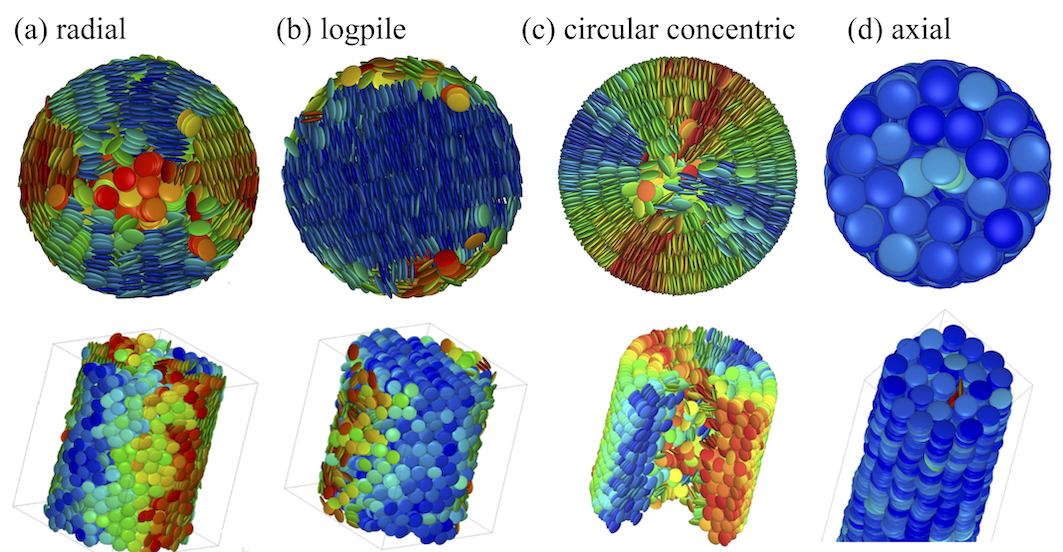}	
		\caption{{\textbf{Monte Carlo simulation top- and side-view snapshots of columnar discotic order in cylindrical nanopores.} In panel (a), (b) textures typical of face-on and in (c), (d) characteristic of edge-on molecular anchoring at the pore wall are depicted, respectively \cite{Zantop2015, Sentker2018}}.}
		\label{fig:MCIllustHydroPhilicPhobic}
	\end{figure*}		
\fi
\subsection{Optical polarimetry} 
The temperature dependent (0.15\,K/min) optical retardation $R(T)$ between perpendicularly polarised ordinary and extraordinary beams can be measured using a high-resolution optical polarimetry technique, see Ref.\,\cite{Kityk2008, Kityk2009} and the Electronic Supporting Information (ESI) for experimental details \footnote{The ESI is available as ancillary file.}. This technique yields direct information about the thermotropic orientational order of DLCs in nanopores. Specifically, a positive $R$ indicates an alignment of the disk-like molecules with their faces in radial direction, and a negative one an orientation of the disk faces parallel to the pore axis \cite{Kityk2008, Calus2014, Kityk2009}.  

\subsection{X-ray diffraction experiments}
In Fig.~\ref{fig:3DRezScattering}(a) the scattering experiment on the liquid-crystal filled nanoporous membranes is schematically depicted, specifically the orientation of the sample rotation axis $\hat{\omega}$ with regard to the AAO membrane and the detector orientation. \revision{To introduce the reader to the 3D reciprocal space mapping and to the expected evolution of the diffraction patterns for the distinct textures upon sample rotation, we show in Fig.~\ref{fig:3DRezScattering}(b) the scattering geometry in analogy to X-ray diffraction texture analysis of fibres \cite{Stribeck2009}. As an example texture the CC state of HAT6 is shown in direct space for a sample rotation $\omega=90\degree$. The cylindrical pore axis $\hat{p}$ (gray rod) agrees with the AAO membrane's surface normal direction and it is perpendicular to the incident beam direction (red arrow) and thus to the incident wavevector $\vec{k_{\rm i}}$ (blue arrow). In the detector plane the wavevector-transfer directions $q_{\rm x}$ and $q_{\rm z}$, the azimuth angle $\chi$ and the orientation of $\hat{p}$ and $\hat{\omega}$ are indicated. Also the plane in direct space encompassing $\hat{p}$ and $\hat{\omega}$ is marked by a transparent blue plane. It is parallel to the detector plane and represents the plane in which translational order is probed in our scattering geometry, i.e. in the $q_{\rm x}-q_{\rm z}$ plane with $q_{y}$=0. The distinct texture discussed here, called (100)$_\parallel$ in the following, has one of the \{100\} plane sets parallel to $\hat{p}$. Thus, the main structural motive in the $\hat{p}$-$\hat{\omega}$ plane in direct space is a hexagonal arrangement of the columns. In reciprocal space this means the occurrence of Bragg intensities at the vertices of the magenta hexagon with a wave vector transfer $q_{\rm{(100)}}=$(0.344 $\pm$ 0.005)\,\AA$^{\rm\rm -1}$ typical of the intercolumnar hexagonal order of HAT6. Because of the concentric circular columnar order about $\hat{p}$ in direct space, the overall Bragg intensity in reciprocal space can be derived from a rotation of the magenta hexagon about the $\hat{p}$ direction which agrees with the direction of $\vec{q}_{\rm x}$, resulting in three yellow Bragg rings centred about the $q_{\rm x}$ axis.  Additional details of the X-ray diffraction analysis can be found in the ESI.}

\revision{
The Bragg condition is fulfilled in those 6 directions (magenta lines), in which in reciprocal space the 3 Bragg intensity rings cut into the Ewald sphere with radius $k_{\rm i}=2\pi/\lambda$, , where $\lambda$ is the X-ray wavelength. This results in a six-fold diffraction pattern on the 2D X-ray detector with Bragg peaks at $\chi_\parallel$=30, 90, 150, 210, 270 and 330\degree, respectively. }
\revision{
The changes in direct and reciprocal space upon sample rotation about $\hat{\omega}$ along with the expected evolution of the detector intensities for the CC $(100)_{\parallel}$ domain upon sample rotation is illustrated in the movie M1 of the ESI \footnote{The movies of the ESI are available at \href{https://doi.org/10.15480/336.2515}{https://doi.org/10.15480/336.2515}.}. Moreover, in the movie M2 of the ESI an illustration of the scattering kinematics is shown for the case that an additional CC domain with one set of $(100)$ planes perpendicular to $\hat{p}$, i.e. a $(100)_{\perp}$ domain is present in direct space. Such a domain is represented by a hexagon rotated by 30\degree~ with respect to the magenta hexagon in Fig.~\ref{fig:3DRezScattering} about $\vec{k_{\rm i}}$. Again the circular concentric columns mean a rotation of this hexagon about $\hat{p}$ in reciprocal space. This results in addition to 2 intensity rings and 2 Bragg peaks in the $\hat{p}$ direction and thus to a six-fold diffraction pattern with Bragg peaks at $\chi_{\perp}$=0, 60, 120, 180, 240 and 300\degree~ for $\hat{\omega}$=90\degree, see the cyan intensities in reciprocal space in the movie M2 of the ESI. }

\revision{
So far, we discussed solely CC structures with regard to their fingerprints in direct and in reciprocal space. Interestingly also for the logpile and radial configuration the main structural motive in direct space is the hexagonal arrangement of the columns \cite{Kityk2014, Zhang2017}, see Fig.~\ref{fig:MCIllustHydroPhilicPhobic}(a,b). Moreover, the rotational symmetry about $\hat{p}$ results for logpile domains with $(100)$ plane sets parallel or perpendicular to $\hat{p}$ to qualitatively identical Bragg intensity rings in reciprocal space as inferred above for the corresponding CC domains. This is illustrated in the movie M2 of the ESI. Thus, distinguishing between those different domains is only possible by a consideration of additional information beyond the bare Bragg peak patterns, like surface anchoring and domain sizes in different directions, as will be discussed in the Results section.  
}

\iftrue
\begin{figure*}
	\centering
	\includegraphics[angle=0,width=1\textwidth]{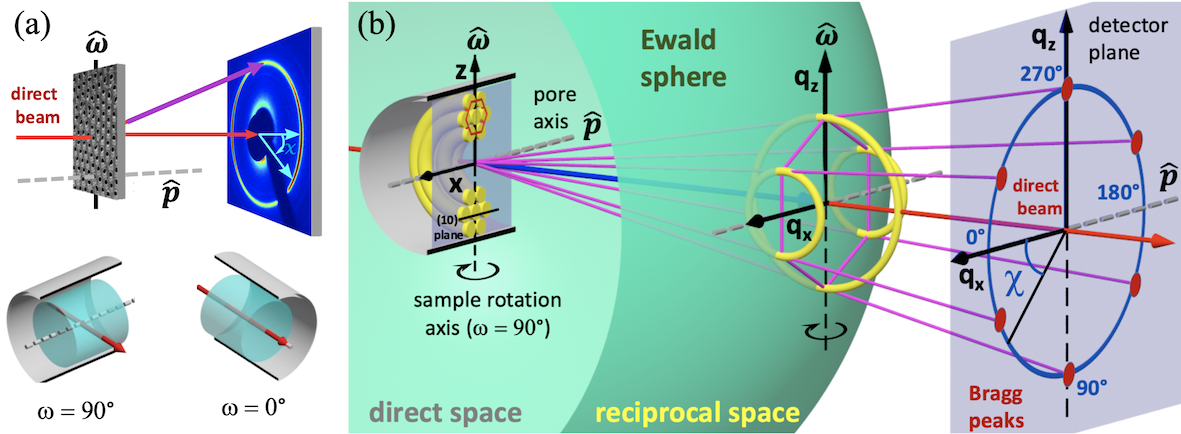}	
	\caption{\textcolor{black}{(a) Schematics of the synchrotron-based X-ray scattering experiment and definition of azimuth angle $\chi$ and sample rotation angle $\omega$ for a configuration with incident X-ray beam parallel or perpendicular to the long cylindrical pore axis direction $\hat{p}$, respectively.(b) 3D X-ray scattering geometry in direct and reciprocal space. Exemplary shown are a discotic columnar liquid crystal with circular concentric ring formation upon confinement in a cylindrical pore in direct space along with the corresponding reciprocal space pattern and the resulting six-fold Bragg diffraction pattern on the detector for a sample rotation $\omega=90\degree$. The sample rotation axis $\hat{\omega}$ (black rod) and the plane encompassing the long pore axis $\hat{p}$ (grey rod) and $\hat{\omega}$ are indicated in direct space (transparent blue plane). It is parallel to the detector plane. The incident wave vector {$\vec{k}_{\rm i}$} (blue arrow), the wavevector-transfer directions $\vec{q}_{\rm x}$ and $\vec{q}_{\rm z}$,  the azimuth angle $\chi$ and the orientation of the direct x-ray beam (red arrow) are also displayed. A detailed discussion of the scattering geometry can be found in the Experimental section~2.3}}
	\label{fig:3DRezScattering}
\end{figure*}
\fi

\subsection{Monte Carlo simulations}
Details on the Monte Carlo simulations can be found in the ESI \cite{swendsenPRL1986,BatesJCP1996, Yan1999,Caprion2003, Earl2005, Lechner2008, caprion2009discotic,Sentker2018}. The main results of these simulations, in particular with regard to the circular concentric columnar state in cylindrical confinement have already been reported in Ref. \cite{Sentker2018}. \revision{Here, they were employed to achieve proper, physics-based visualisations of the distinct confined textures.}

\section{Results and discussion}

\subsection{Self-assembly in hydrophilic nanopores with face-on discotic wall anchoring}
\subsubsection{X-ray diffraction and translational order}
\iftrue
\begin{figure*}[htbp]
    \centering
    \includegraphics[width=0.95\textwidth]{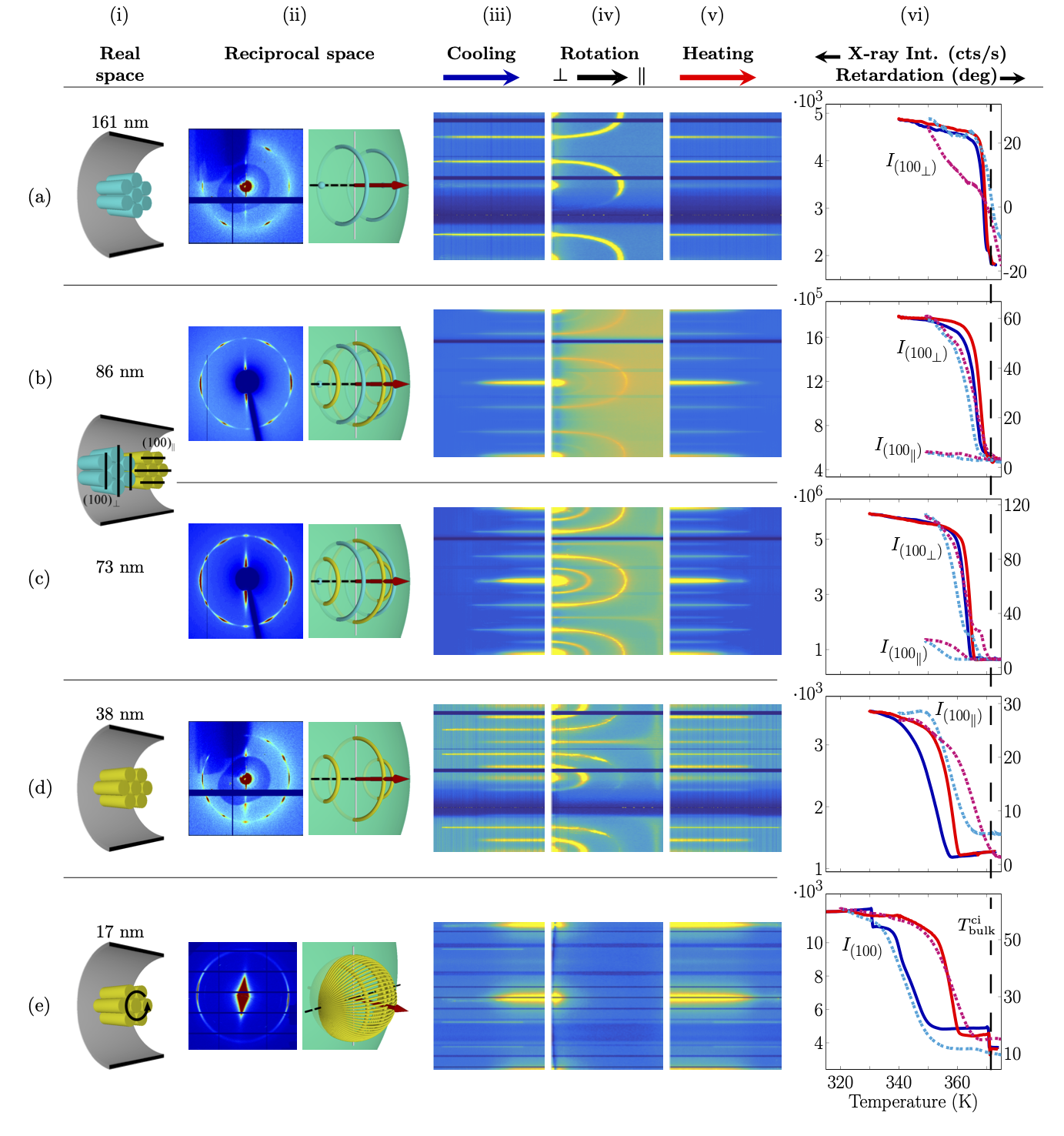}
    \caption{\textbf{Self-assembly of discotic molecules in hydrophilic cylindrical nanopores (face-on molecular anchoring at the pore wall) as inferred from x-ray diffraction and birefringence measurements.}  (i) Columnar order in real space, (ii) measured x-ray diffraction pattern and illustration of the corresponding scattering geometry and Bragg intensities in reciprocal space encompassing the Ewald sphere. (iii) Bragg intensities as a function of temperature for a cooling ($\omega =$ 85$^{\rm\circ}$), (iv) sample rotation ($\omega =$ 90$^{\rm\circ}\rightarrow$ 0$^{\rm\circ}$) and (v) heating ($\omega =$ 85$^{\rm\circ}$) scan extracted for each azimuth angle $\chi$. See definition of azimuth angle and axis labelling in Fig.\,\ref{fig:3DRezScattering}. Temperature dependent retardation experiments (solid curves) in comparison with $T$-dependent Bragg intensities (broken lines) typical of the hexagonal columnar translational order. The dashed line indicates the bulk isotropic-columnar transition temperature $T^{\rm\rm bulk}_{\rm ci}$. \textcolor{black}{The samples were cooled down in the x-ray experiments from 375~K to (a-c) 350~K, (d) 340K and (e) 320~K. The x-ray data were taken at a wavelength of (a,d) 0.177~\AA, (b,c) 0.496~\AA~and (e) 0.729~\AA.}}
    \label{fig:SummaryAAON}
\end{figure*}  
\fi	
In Fig.\,\ref{fig:SummaryAAON} we present X-ray diffraction measurements for native, hydrophilic pore walls and selected pore diameters. The 2D diffraction patterns were recorded after cooling from the isotropic phase ($T\,>\,T_{\rm\text{bulk}}\,=$\,371.1\,K) down to the liquid crystalline phase (see minimal temperatures depending on pore diameter in the figure caption) \revision{with a cooling rate of 1 K/min at $\omega=85\degree$}. 

We start discussing the diffraction experiments performed on the largest pore diameter, $d =$ 161\,nm. We observe a six-fold Bragg reflection pattern at  $q_{\rm{(100)}}=$(0.344 $\pm$ 0.005)\,\AA$^{\rm\rm -1}$ typical of the intercolumnar hexagonal order of HAT6, see Fig.~\ref{fig:SummaryAAON}a(ii). The pattern is aligned with one of the 100 Bragg peaks $\parallel \hat{p}$, i.e. the \{100\} plane set $\perp \hat{p}$. This corresponds to the $(100)_{\perp}$ domain, see Fig.\,\ref{fig:SummaryAAON}a(i) and b(i). Upon $\omega$ sample rotation, i.e. a rotation from a configuration with $\vec{k_{\rm i}}\parallel\hat{p}$ to $\vec{k_{\rm i}}\perp\hat{p}$ the Ewald sphere cuts under an increasing angle into this reciprocal space pattern. This is in excellent agreement with the intensity contour plots of the $\omega$-scans, see Fig.\,\ref{fig:SummaryAAON}a(iv). Initially one observes six Bragg peaks in azimuthal direction. Upon $\omega$ rotation the Bragg condition of the {(100)}$_{\perp}$ planes is not fulfilled anymore. They vanish, whereas the $\chi$ position of the other plane sets starts to move towards the polar and equatorial directions, respectively, in perfect agreement with a rotation of the ring structures depicted in Fig.\,\ref{fig:SummaryAAON}a(ii). See also the evolution of the cyan diffraction intensities representing the $(100)_{\perp}$ domain upon sample rotation in the movie M2 in the ESI. 

As discussed in the Experimental section 2.3 radial and logpile structures have indistinguishable footprints in reciprocal space. However, in case of the logpile structure the domain size should be perpendicular to the long pore axis on the order of the pore diameter \cite{Zhang2017}. Whereas in the case of radial columns this perpendicular coherence length, $\xi_{\perp}$ is expected to be much smaller. Here we determined a value of $\xi_{\perp}$=153\,nm. This value is only slightly smaller than the nominal pore diameter and thus indicating a logpile arrangement of the molecules.
	
Upon decreasing the diameter down to 73\,nm the pattern changes from a six-fold to a twelve-fold pattern, see Figs.~\ref{fig:SummaryAAON}b(ii) and c(ii). This can be understood by the appearance of a second hexagonal orientational domain, i.e. the $(100)_{\parallel}$ orientational domain, see Fig.\,\ref{fig:SummaryAAON}b(i). The $\omega$-rotation contour plots (Figs.~\ref{fig:SummaryAAON}b(iv) and c(iv)) are compatible with the existence of these two domains, see also the movie M2 in the ESI. The correlation length is also for this pore size on the order of the pore diameter. A logpile structure is established with a $\xi_{\perp}$ of 65~nm, see Fig.~\ref{fig:Coherencelength}, where $\xi_{\perp}$ is plotted as a function of pore size.

Interestingly, upon further decreasing $d$, we see a transition from a twelve-fold pattern to a six-fold pattern, see Fig.~\ref{fig:SummaryAAON}d(ii). \footnote{Note that we observe a second set of Bragg reflections at $\chi$s typical of the $(100)_{\perp}$ domain for the sample with 38\,nm pore size. It appears at temperatures much higher than for the confined liquid crystal, close to the bulk transition, and exhibits very sharp Bragg reflections indicating large coherence lengths. Therefore, we attribute these reflections to a textured bulk film growing with one (100) plane set parallel to the AAO membrane surface, rather than tracing them to a $(100)_{\perp}$ domain in pore space. However, we cannot entirely exclude the latter possibility.} This means that only one of the orientational domains survives, i.e. the $(100)_{\parallel}$ domain (see Fig.~\ref{fig:SummaryAAON}d(i). The correlation lengths and the $\omega$-Bragg intensity contour map are again compatible with the logpile structure, see Fig.~\ref{fig:Coherencelength}.
\iftrue
\begin{figure}[tbp]
	\centering
	\includegraphics[width=0.45\textwidth]{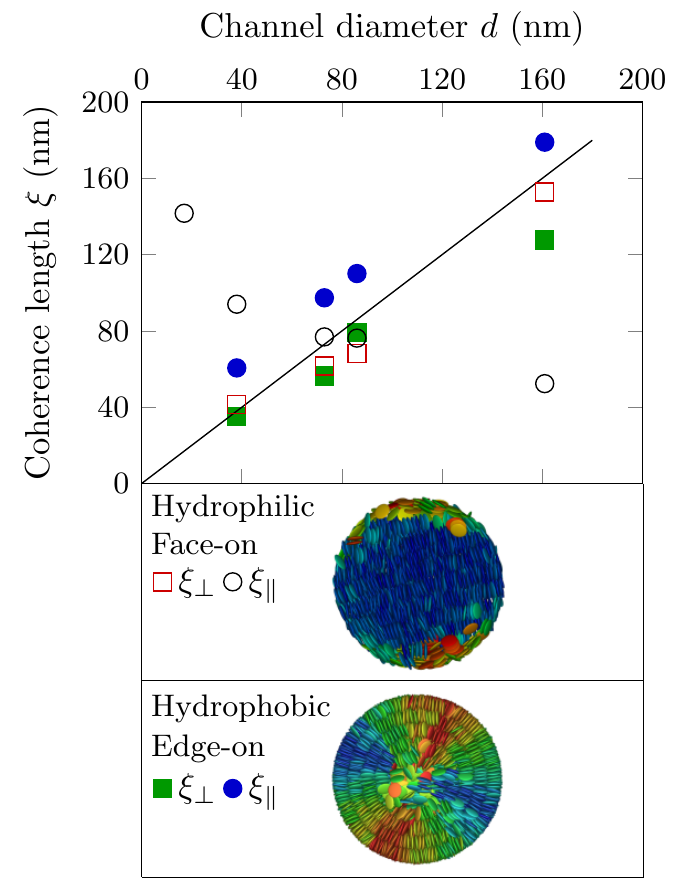}		
	\caption{\textbf{Coherence lengths $\xi$ of translational columnar order parallel and perpendicular to the long pore axis as a function of channel diameter and hydrophobicity of the pore walls.} Shown are $\xi_{\perp/\parallel}$ for the logpile configuration in hydrophilic and the circular concentric configuration in hydrophobic pores. The straight line indicates $\xi$ = pore diameter $d$.}
	\label{fig:Coherencelength}
\end{figure}
\fi	
At the smallest pore diameter a diffraction pattern with two strong intensity halos in the polar directions is observed, see Fig.~\ref{fig:SummaryAAON}e(ii). It can be understood by a logpile structure along with a randomisation of the [hk0] directions with regard to the long pore axis, as shown in Fig~\ref{fig:SummaryAAON}e(i). This results in a densification of the Bragg intensities towards the equatorial directions as shown in Fig.\,\ref{fig:SummaryAAON}d(ii) \cite{Zhang2017}. \revision{Hence, we conclude that for all pore-sizes studied we observe a logpile structure in the case of face-on anchoring at the pore wall.} 

Additionally, to the diffraction patterns discussed above, X-ray diffraction patterns upon cooling from, and subsequent heating back into the isotropic phase were recorded. In Fig.\,\ref{fig:SummaryAAON} the Bragg intensity at a wavevector transfer typical of the columnar self-assembly is plotted as a function of azimuthal angle (vertical axis) and temperature (horizontal axis) for cooling and heating in column (iii) and (v), respectively. Upon cooling/heating one can observe the increase/vanishing of intensities characteristic of the isotropic-to-columnar and columnar-to-isotropic transition and hence the specific occurrence of the Bragg intensities characteristic of the $(100)_{\parallel}$ and $(100)_{\perp}$ orientational domains as a function of temperature.

\subsubsection{Optical birefringence and collective orientational order}

It is interesting to compare this information on the translational self-assembly with the information gained from our temperature-dependent optical retardation experiments regarding the evolution of the orientational order in confinement. In Fig.\,\ref{fig:SummaryAAON}(vi) we plot the intensities of different Bragg peak sets corresponding to the characteristic domain states upon cooling (blue) and heating (red) in comparison with the optical retardation. A decrease (increase) of $R$ with decreasing (increasing) temperature indicates an averaged orientation of the director $\hat{n}$ parallel (perpendicular) to the pore axis $\hat{p}$ \cite{Kityk2014}. The retardation for all pore diameters studied increases upon cooling indicating a collective orientation of the in-plane disc direction of the molecules along the long channel axis of the pores. This is in perfect agreement with the conclusion drawn from the X-ray diffraction experiments regarding the translational logpile order for all pore sizes studied. Moreover, the retardation behaves analogously to the Bragg intensity upon cooling and heating. In contrast to the abrupt bulk transition of first order, the transition evolves continuously. A substantial hysteresis between heating and cooling, which significantly increases upon decreasing pore size is present. Additionally, the temperature evolution of the x-ray intensities indicates that the $(100)_{\parallel}$ domain appear much weaker at lower temperatures compared to the $(100)_{\perp}$ orientation in the case of a coexistence of these domains. This is particularly obvious for the 73\,nm and 86\,nm pores, see Fig.~\ref{fig:SummaryAAON}c(vi) and b(vi). Finally, for the smallest pore diameter ($d =$ 17\,nm) we get an excellent agreement for the evolution of the collective orientational (retardation) and translational hexagonal columnar order (100) Bragg intensity, see Fig.~\ref{fig:SummaryAAON}e(vi). In both cases a strong supercooling and a large hysteresis between cooling and heating is observed. Complementary data sets can be found in the supplementary. They are consistent as a function of $R$ with the results discussed here for a representative set of measurements. 

Overall, our findings are in excellent agreement with the observations and interpretations of Zhang et al. \cite{Zhang2017}. Nevertheless our experiments yield complementary detailed information about the temperature dependence.  We also observe peculiar changes of the orientational domain texture with regard to the long pore axis as a function of $R$. It has been established for crystallisation in unidirectional confinement that extreme textures can arise resulting from the competition of fast growing crystallisation along the long channel axis, i.e. Bridgman growth in narrow capillaries, with crystallisation constraints and/or molecular anchoring at the pore walls \cite{Henschel2007, Henschel2009, Knorr2009}. The $(100)_{\perp}$ orientation can presumably be traced to Bridgman growth, if one assumes that the $<100>$ direction is a fast growing mode and thus is aligned along $\hat{p}$. As discussed in detail by Zhang et al. for the smaller pore diameters an increasing contribution of splay deformations at the pore walls sets in. These defect structures, also found in MC simulation studies \cite{Zantop2015}, can be more easily established in the $(100)_{\parallel}$ orientation. Thus we conclude that the peculiar changes in the orientational domain structures result from transitions from a Bridgman-growth determined to splay-deformation determined orientational states as a function of reducing pore diameter.

To gain additional information on this confinement-guided texturing and the domain sizes, we determined also the coherence lengths along the long pore axis $\xi_{\parallel}$ from the detected Bragg peaks, see Fig.~\ref{fig:Coherencelength}. \revision{In contrast to $\xi_{\perp}$ it decreases with decreasing $d$. This interesting behaviour is supportive of the Bridgman growth mechanism and the splay-induced domain selection and growth along the long channel axis. Both are increasingly effective with increasing anisotropy of the confining geometry, i.e. with decreasing $d$ \cite{Henschel2009}.}

\subsection{Self-assembly in hydrophobic nanopores with edge-on discotic wall anchoring}
\subsubsection{X-ray diffraction and translational order}
\iftrue
\begin{figure*}[htbp]
    \centering
   \includegraphics[width=0.95\textwidth]{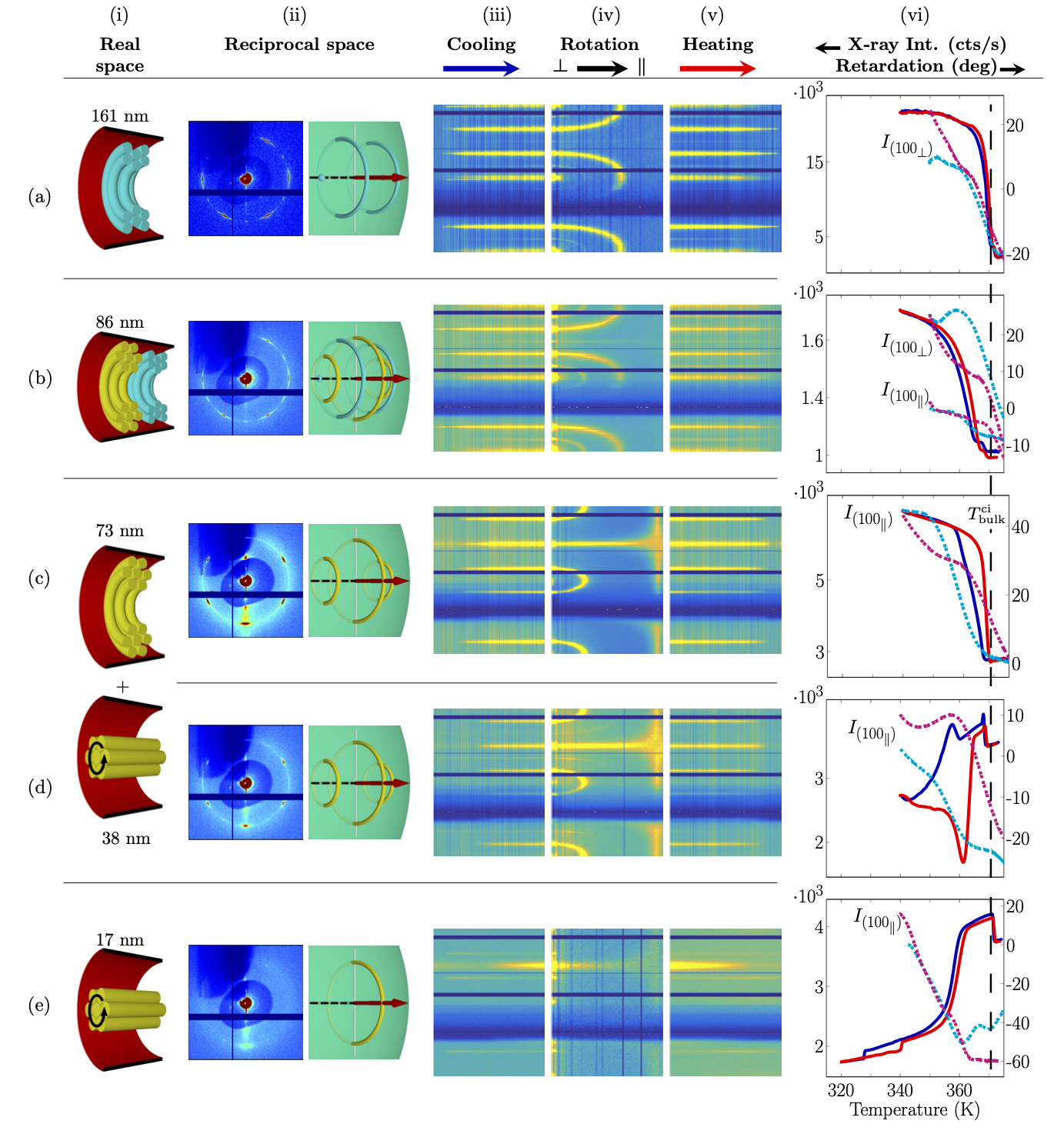} 
    \caption{\textbf{Self-assembly of discotic molecules in hydrophobic cylindrical nanopores (edge-on molecular anchoring at the pore wall) as inferred from x-ray diffraction and birefringence measurements.} (i) Columnar order in real space, (ii) measured x-ray diffraction pattern and illustration of the corresponding scattering geometry and Bragg intensities in reciprocal space encompassing the Ewald sphere. (iii) Bragg intensities as a function of temperature for a cooling ($\omega =$ 85$^{\rm\circ}$), (iv) sample rotation ($\omega =$ 90$^{\rm\circ}\rightarrow$ 0$^{\rm\circ}$) and (v) heating ($\omega =$ 85$^{\rm\circ}$) scan extracted for each azimuth angle $\chi$. See definition of azimuth angle and axis labeling in Fig.\,\ref{fig:3DRezScattering}. (vi) Temperature dependent retardation experiments (solid curves) in comparison with $T$-dependent Bragg intensities (broken lines) typical of the hexagonal columnar translational order. The dashed line indicates the bulk isotropic-columnar transition temperature $T^{\rm\rm bulk}_{\rm ci}$. \revision{The samples were cooled down in the X-ray experiments from 375~K to (a, b) 350~K and (c-e) 340~K. All X-ray data are taken at a wavelength of 0.177~\AA.}}
    \label{fig:SummaryAAOODPA}
\end{figure*} 
\fi

In Fig.\,\ref{fig:SummaryAAOODPA} we present X-ray diffraction measurements for a selected set of pore diameters for surface-grafted, hydrophobic pore walls. Again the 2D diffraction patterns were recorded after cooling from the isotropic phase ($T\,>\,T_{\rm\text{Bulk}}\,=$\,371.1\,K) at $\omega=85\degree$, i.e. $\vec{k_{\rm i}}\perp\hat{p}$ with a cooling rate of 1\,K/min. For large pore diameters, we observe again a transition from a six-fold to a twelve-fold diffraction pattern, see Fig.~\ref{fig:SummaryAAOODPA}a-d(i). The $\omega$-intensity contour plots once more result from two orientational domains with regard to $\hat{p}$, a $(100)_{\parallel}$ and $(100)_{\perp}$ domain, as illustrated in Fig.\,\ref{fig:SummaryAAOODPA}b(ii). However, given the edge-on surface anchoring they do not result from logpile structures, but from hexagonal arranged circular concentric bent columnar structures as illustrated in Figs.\,\ref{fig:MCIllustHydroPhilicPhobic}c and \ref{fig:SummaryAAOODPA}a-d(i) \cite{Zhang2012, Zhang2014, Zhang2019}. See also movie M4, where virtual and real scattering patterns are shown in direct comparison as a function of $\omega$ rotation for $d$=86~nm.

For small pore diameters down to 38\,nm only a six-fold pattern remains (Figs~\ref{fig:SummaryAAOODPA}c-d(ii) and the movies M1 and M5 in the ESI), compatible with the existence of the $(100)_{\perp}$ orientation. More importantly, the $\omega$-contour plots evidence that starting with $d=73$ nm for $\omega=0 \degree$~a Bragg peak in the polar directions occur ($\chi=\pm90\degree$), see Figs~\ref{fig:SummaryAAOODPA}c-e(iv). These partly result from an additional domain state with columns parallel $\hat{p}$, i.e. axial columns. It results in one ring in the $k_{\rm i}$-$\hat{\omega}$ plane at $q_{(100)}$ in reciprocal space and thus to the appearance of two Bragg peaks at $\chi$=90\degree~ and 270\degree~, see Fig.~\ref{fig:SummaryAAOODPA}e(ii) and the movies M3 and M6 in the ESI. For the smallest pore diameter, $d =$ 17\,nm the hexagonal patterns are only marginally detectable, see Fig.\,\ref{fig:SummaryAAOODPA}e(ii). Only polar intensities are present. An almost purely axial columnar arrangement is present. This molecular stacking along the pore axis is confirmed by the position of the intracolumnar disc-disc peak at $q_{\rm dd}= $(1.763 $\pm$ 0.018)\,\AA$^{\rm\rm-1}$ on the horizontal axis (not shown).

Thus, there is a textural transition between circular concentric at large and axial columnar order at small pore diameters. What drives this transition? As suggested by Ungar et al. \cite{Zhang2015, Zhang2019} it can be rationalised by the competition of the bending energy typical of the CC configuration and the energetic cost for the distortion of the 2D columnar lattice at the cylindrical pore wall in the case of the axial columnar state. Whereas the former scales linearly with the pore diameter $d$, the latter scales with $\log{d}$. Thus, below a critical radius $d_c$ the axial state should be favoured compared to the radial one, in agreement with our observation. We find $d_{c} \sim 70$~nm.

Interestingly, this transition and thus pure axial orientation of neat HAT6 in cylindrical nanopores has (to the best of our knowledge) never been observed before. Specifically, chemical pore-surface grafting with silanes, i.e. silanisation, to favour edge-on anchoring did not result in this configuration \cite{Kityk2014, Zhang2015}. Presumably, this dichotomy can be traced to a more robust and homogeneous hydrophobic surface coating by the phosphonic functionalisation in comparison to silanisation. In fact, surface-sensitive X-ray diffraction at planar surfaces indicates that silanisation works well on hydroxylated silica surfaces \cite{Steinruck2014,Khassanov2015}. By contrast, on planar aluminium oxide surfaces, as employed here, phosphonic acid linking gives more ordered self-assembled monolayers than silanisation \cite{Klauk2007, Khassanov2015}. 

\subsubsection{Optical birefringence and collective orientational order}
\iftrue
\begin{figure}
	\centering
	\includegraphics[width=1\columnwidth]{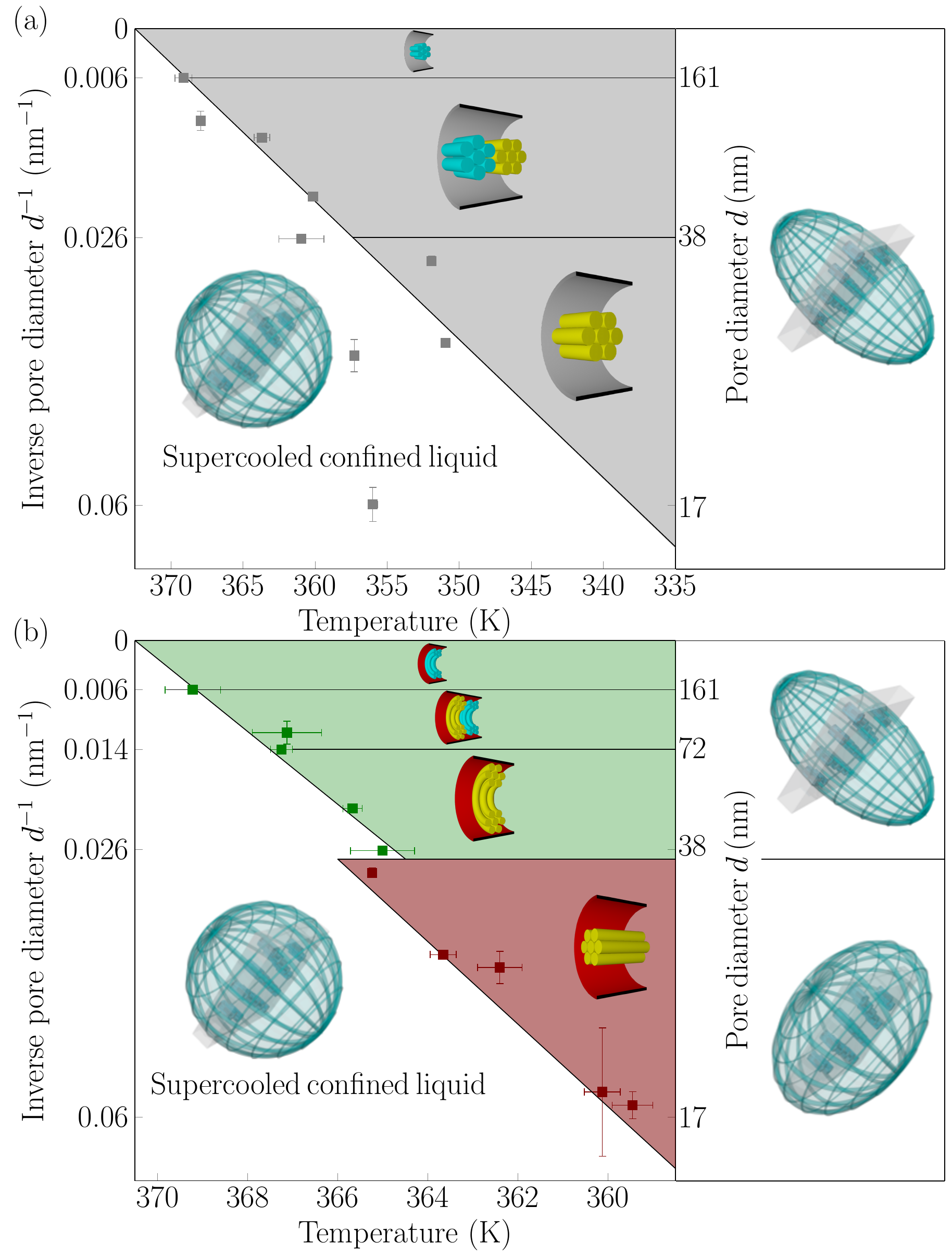}		
	\caption{\textbf{Phase diagram and effective optical anisotropy of liquid crystal-infused membranes.} The symbols represent the pore-size dependent columnar-to-isotropic phase transition temperatures $T_\text{col-iso}$ for membranes with (a) hydrophilic and (b) hydrophobic cylindrical nanopores as determined from optical birefringence experiments. The solid diagonal lines separating the supercooled confined liquid state and the confined columnar phases are $1/d$ fits of $T_\text{col-iso}$. The insets represent sketches of the columnar order in the single nanopores along with the resulting effective isotropy in the liquid and positive or negative optical optical birefringence in the liquid-crystalline state, i.e. spherical, prolate or oblate indicatrices, respectively.}
	\label{fig:HAT6PhaseDiagramMetaOptics}
\end{figure}
\fi

Again it is interesting to compare these diffraction results to the optical experiments. The qualitative behaviour is quite similar to the observations in the hydrophilic pores. A rather gradual evolution of the Bragg intensities compared to the sharp bulk transition. Once more the different domains occur and vanish at slightly different temperatures. Additionally, there seems to be a competition between the different orientational domains as a function of cooling/heating history. This competition occurs for pores in the size range between 73\,nm and 161\,nm only between positive birefringent contributions, i.e. between the $(100)_{\parallel}$ and $(100)_{\perp}$ domains, see Fig.~\ref{fig:SummaryAAOODPA}a-c(ii). In case of 38\,nm pores competing axial and circular concentric order contribute to $R(T)$, and thus positive and negative contributions respectively, are present. Consequently, the optical birefringence shows a non-monotonous behaviour both upon cooling and heating, see Fig.~\ref{fig:SummaryAAOODPA}d(vi). In particular, the birefringence experiments show that upon cooling first radial alignment with positive $R$-contributions is established. However upon cooling an additional axial arrangement giving negative contributions evolves. As shown in Fig.~\ref{fig:SummaryAAOODPA}e(vi), only for the smallest pore diameter, the birefringence continuously drops and thus indicates an orientation of the in-plane disc orientation $\perp \hat{p}$ in excellent agreement with the formation of an almost purely axially aligned columnar state inferred from the diffraction experiments.
\iftrue
\begin{figure}[h]
	\centering
	\includegraphics[width=1\columnwidth]{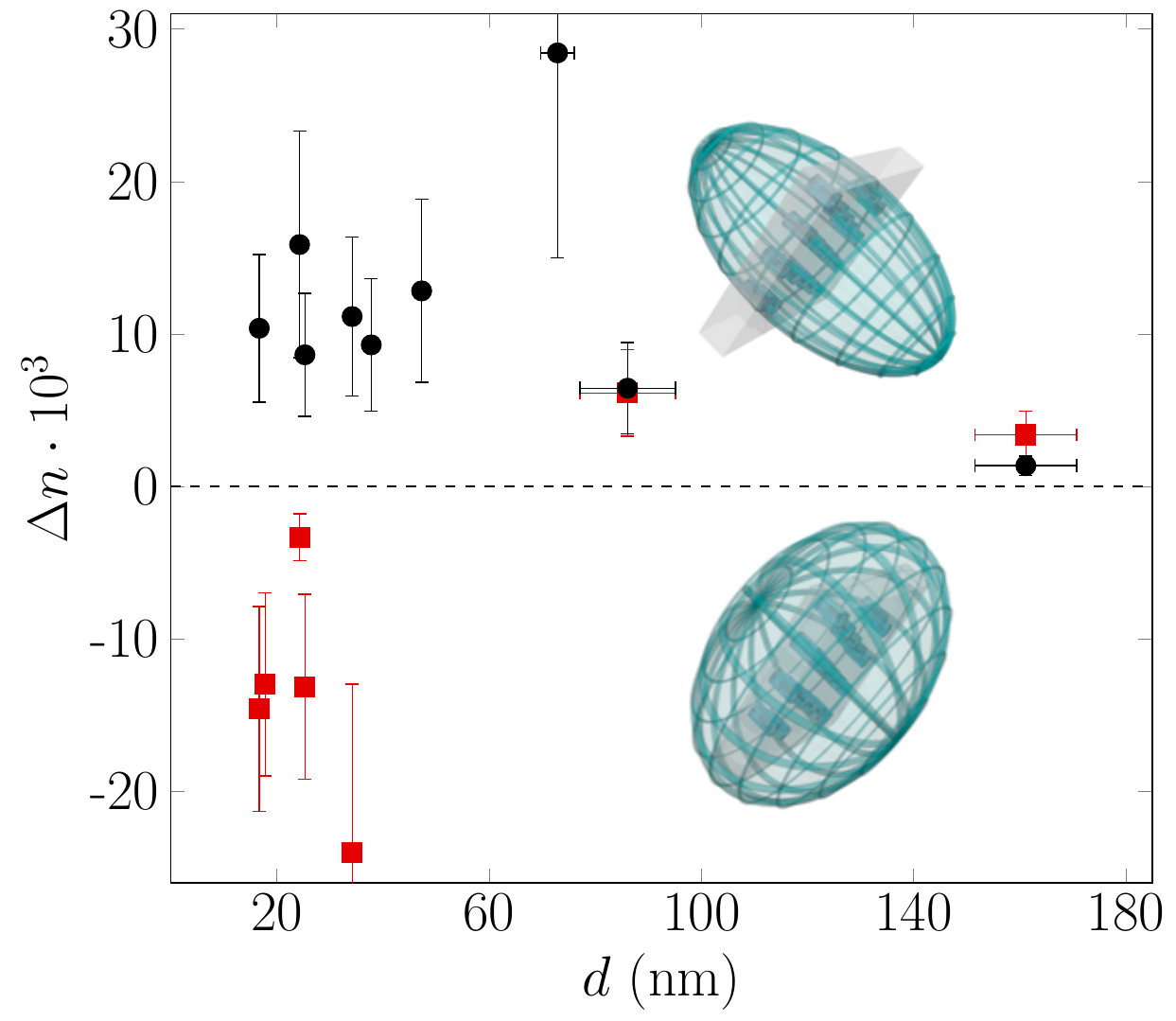}		
	\caption{\textbf{Effective birefringence.} Effective birefringence of liquid crystal-infused anodic aluminium oxide as a function of pore diameters for hydrophilic (black circles) and hydrophobic (red squares) nanopores in the liquid crystalline phase.}
	\label{fig:birefringence}
\end{figure}
\fi

\iftrue
\begin{figure}[htbp]
	\centering
	\includegraphics[width=0.5\textwidth]{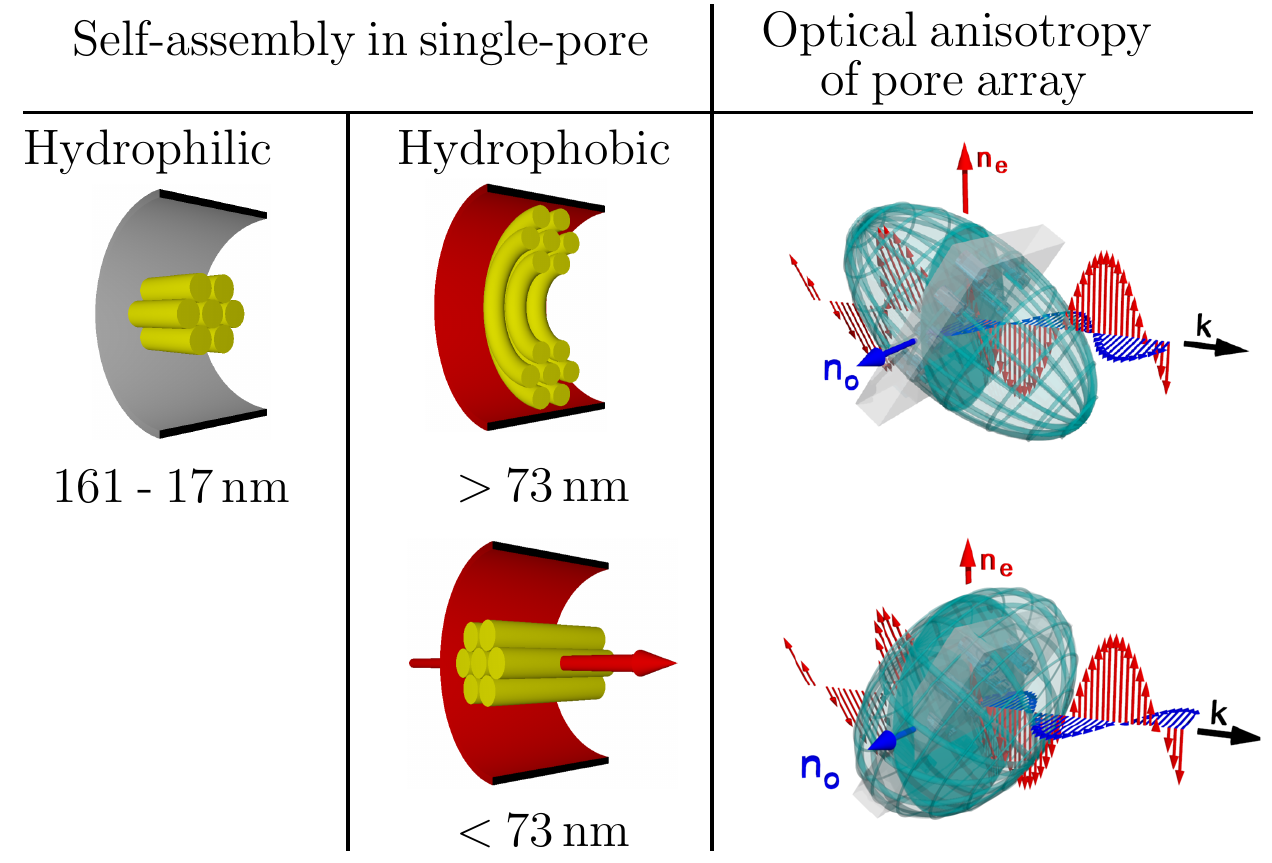}	
	\caption{\textbf{Confinement-controlled self-assembly and resulting effective optical anisotropy.} Pore-size and surface-hydrophilicity dependent self-assembly on the single-pore scale and resulting optical anisotropy of the nanopore array.}
	\label{fig:CompBirefringencePositiveNegative}
\end{figure}
\fi
\iftrue
\begin{figure*}[]
	\centering
	\includegraphics[width=0.7\textwidth]{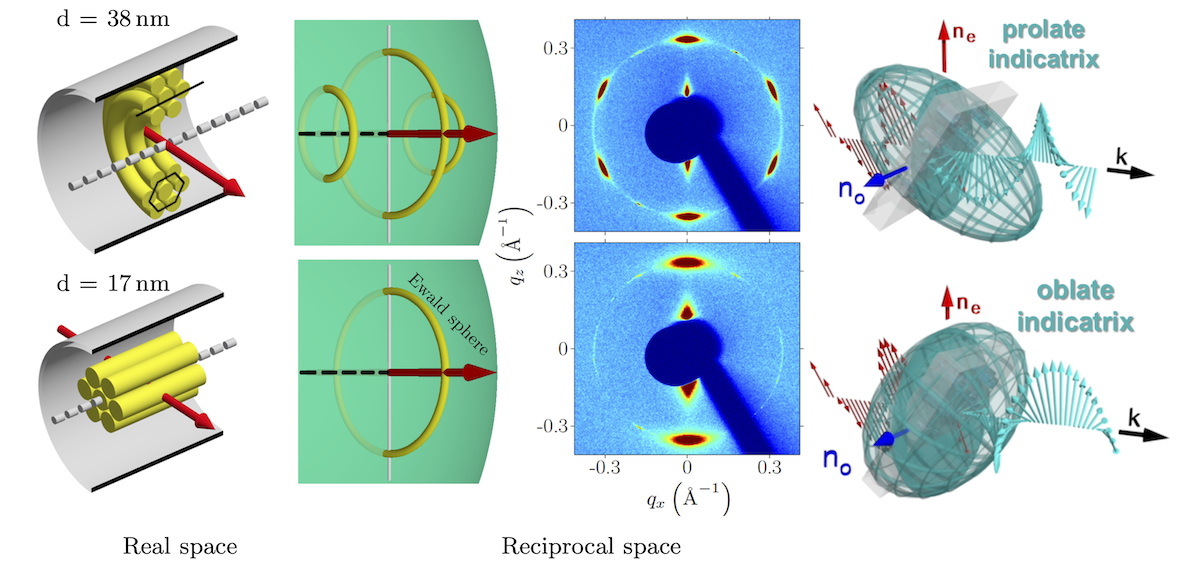}	
	\caption{\textbf{Distinct columnar textures on the single-pore scale and the resulting polarisation state of light transmitted through the pore array.} Self-assembly of radial aligned rings and axial aligned columns of disk-like molecules on the single-pore scale, as evidenced by 2 D X-ray diffraction, results in an either prolate or oblate ellipsoid of refractive indices (indicatrix) aligned to the pore axis direction. Thus, linear polarised light is split up by a parallel array of such pores into two beams with perpendicular polarisations and distinct propagation speeds. Their relative phase shift after passing the birefringent pore array is positive or negative. Thus, they superimpose to left- and right-hand elliptically polarised light, respectively, for the specific beam and membrane geometry of our experiments.}
	\label{fig:RadialAxialXRayBirefringence}
\end{figure*}
\fi

Encouraged by findings in MC simulations \cite{Zantop2015, Sentker2018} we rationalise this complex thermal history as follows. Dictated by immobilisation at the pore walls and the large curvature radius, and thus smaller excess energy for circular ring formation, the formation of circular concentric structures starts at the pore walls. Then it propagates as a function of cooling from the pore periphery to the pore centre. Thus from large columnar bent radii to small ones. This results in the monotonic behaviour of the $R(T)$ curves \cite{Kityk2014}. Interestingly, upon heating significant drops in $R(T)$ indicate that part of the circular concentric configurations transform to the axial configuration just before the complete melting of the columnar state, e.g., see $R(T)$ for $d$=38\,nm in Fig.~\ref{fig:SummaryAAOODPA}d(vi). This suggest that for these pore sizes only small energy differences between the distinct textures exist.
	
\subsection{Thermotropic phase behaviour}
The $R(T)$ curves in column (vi) of Figs.\,\ref{fig:SummaryAAON},\ref{fig:SummaryAAOODPA} and in Supplementary Fig.\,S3 can be used to determine the isotropic-columnar transition temperatures by taking the maximum of the derivative $dR/dT$ \cite{Kityk2014}. This results in the phase diagram shown in Fig.\,\ref{fig:HAT6PhaseDiagramMetaOptics}, where the phase transition temperatures $T^{\rm\rm conf}_{\rm iso-col}$ are plotted versus the inverse pore diameter $d$ for the hydrophilic and hydrophobic pores as determined upon heating. The characteristic textures are illustrated by insets. As indicated by the good agreement with the solid lines representing $1/d$ fits of the phase boundaries, the phase transition shifts follow the Gibbs-Thomson scaling for a phase transition of first order in confined geometry, i.e. $T^{\rm\rm conf}_{\rm iso-col} = T^{\rm\rm bulk}_{\rm iso-col}-K_{\rm GT}\cdot d^{\rm\rm-1}$, where $K^{\rm GT}$ depends on the shape of the phase boundary, thermodynamic parameters of the low temperature phase, specifically the transition enthalpy $\Delta H$, as well as interfacial energies \cite{Webber2007}. For the hydrophilic case $K^{\rm GT}_{\rm logpile}$=517.1$\pm$ 49.1 nm\, K, for the hydrophobic case $K^{\rm GT}_{\rm CC}$= 262.2$\pm$18.0 nm\,K and $K^{\rm GT}_{\rm axial}=$197.2$\pm$15.0 nm\, K. Thus, the confinement induces a much stronger supercooling of the isotropic-to-columnar transition for the hydrophilic in comparison to the hydrophobic case and for the hydrophobic case distinct supercooling parameters for the circular concentric and axial formation are observed. These temperature shifts are within the error bars in agreement with calorimetry and dielectric spectroscopy experiments recently reported \cite{Yildirim2019}. Assuming cylindrical phase boundaries and transition enthalpies in agreement with the bulk value, it is possible to attribute these differences to distinct interfacial tensions $\sigma$ between the columnar phase and the isotropic liquid for the different textures with the Gibbs-Thomson equation \cite{Yildirim2019}: $\sigma_\text{logpile} =$ (3.76 $\pm$ 0.35)\,mN/m, $\sigma_\text{axial} =$ (1.43 $\pm$ 0.1)\,mN/m and $\sigma_\text{CC} =$ (1.91 $\pm$ 0.12)\,mN/m \cite{Yildirim2019}. 

However, it is important to note that given the quite inhomogeneous low-temperature state, the application of classical thermodynamics of homogeneous phases separated by well-defined phase boundaries is questionable. The many grain boundaries in the logpile configuration and disordered regions at the pore walls as well as the elastic excess energies due to columnar bending or deformation of the columnar lattice will result in relative changes of the chemical potentials of the low-temperature and the isotropic liquid state, in particular changes in the transition enthalpies \cite{Yildirim2019}, which additionally, beyond the Gibbs-Thomson considerations, contribute to the phase transition shifts \cite{Huber1999, Schaefer2008}.
\revision{Note that we found in exploratory experiments no hints for a cooling/heating rate dependence of the phase transition shifts, type of transition and hysteresis, see Fig.~S2 of the ESI for $R(T)$ measurements with cooling/heating rates with 0.075, 0.15 and 0.3K/min, respectively, for hydrophobic pores ($d$=17~nm).
} 

\subsection{Effective optics of the liquid-crystal-infused nanoporous solids}
In the previous section the polarimetry measurements were devised to infer the collective orientational order of HAT6 in the confined state. By contrast, we are going to analyse here these results with respect to the effective optical properties of the resulting liquid-crystal infused nanoporous solids. The birefringent properties of the resulting hybrid materials $\Delta n(d)$ as a function of pore surface grafting and thus molecular anchoring are analysed.

Despite the fact that the base materials of the pore walls are optical isotropic the anisotropic, collective arrangement of parallel cylindrical pores results in a small positive geometric birefringence \cite{Saito1989, Kityk2009}. This form birefringence is on the order of $\Delta n \sim$ 0.01 for the empty monolithic nanoporous solids \cite{Gong2011}. Upon filling the pores with the isotropic liquid this small form anisotropy is slightly reduced, because of the reduction in the refractive index difference between pore wall and pore filling. In the following the effective optical anisotropy in terms of the optical birefringence $\Delta n$ of the hybrid systems is analysed. 

Upon isotropic-to-columnar self-assembly in the hydrophobic pore space sizeable positive and negative retardations are observable at low temperature, see column (vi) in Fig.\,\ref{fig:SummaryAAON} and \ref{fig:SummaryAAOODPA}. As is illustrated in the phase diagram shown in Fig.\,\ref{fig:HAT6PhaseDiagramMetaOptics} by prolate and oblate optical indicatrices the complex, hydrophilicity-dependent self-assembly behaviour results in tailorable optical anisotropy. Whereas for the hydrophilic pores always positive optical anisotropy is observable for the pore array, the optical birefringence at low temperature, in the columnar phase, is negative for small pore diameter and positive for large pore diameters in the hydrophobic case. Thus, the optical anisotropy of the metamaterial is switched between positive and negative as expected by the distinct change in the collective arrangement of the mesogens from circular concentric to axial. A more quantitative analysis of the birefringence behaviour can be found in Fig.\,\ref{fig:birefringence}, where we plot $\Delta n$ versus pore size as determined by an effective medium analysis considering the Bruggeman equation, see the ESI. As expected the logpile configuration in the hydrophilic pore space results solely in positive birefringent properties. \textcolor{black}{The largest positive birefringence of 0.03 we inferred for the 73~nm pores, however, with a large error bar because of uncertainties in the porosity of the sample batch. Otherwise, the birefringence increases with decreasing pore size up to values of $\Delta n \sim$ 0.015.} This optical effect can be directly correlated to the increased hexagonal order determined by X-ray diffraction and the resulting improved collective orientational order, see Fig.\,\ref{fig:Coherencelength}. In the hydrophobic case only samples with pure axial or circular concentric, hence pure negative or positive optical anisotropy are considered. The strength of $\Delta n$ in hydrophobic large pores basically equals that of the hydrophilic pores of same size. \textcolor{black}{Upon transition to the pure axial configuration the birefringence switches to negative values. Its absolute value is considerably larger than in the circular concentric case.} 

We summarise the effective optical properties in terms of pore-size dependence and hydrophilicity in Fig.\,\ref{fig:CompBirefringencePositiveNegative}. By going to smaller pores the optical anisotropy can be changed in controlled manner. \textit{i.e.} adapted from positive to negative optical birefringence. Moreover, the onset of sizeable positive and negative birefringence can be systematically tuned by up to 35 K for the hydrophilic and over 15~K for the hydrophobic case by a proper choice of the pore size compared to the fixed bulk value. 

It is interesting to note that the radial-to-axial transition results not only in a change of the optical anisotropy. For our beam geometry and the observed retardations it also means that for the textures with collective radial orientations left-handed elliptically polarised, and for the axial state right-handed elliptically polarised leaves the pore array, respectively, as illustrated in Fig.\,\ref{fig:RadialAxialXRayBirefringence}. Of course, this does not mean that we introduce chirality into the system, rather this effect solely results from the sign change in the phase difference between the ordinary and extraordinary beams.

\section{Conclusions}
Using synchrotron-based 3D reciprocal space mapping, high-resolution optical birefringence measurements and Monte Carlo simulations we investigated the self-assembly behaviour and the resulting optical properties of a thermotropic discotic liquid crystal (HAT6) embedded in nanoporous anodic aluminium oxide as a function of pore size and pore surface functionalization. Independent of pore size cylindrical hydrophilic pores with planar discotic anchoring lead solely to columns radially aligned to the cylindrical pores and thus in a positive birefringence. By contrast, for hydrophobic pores with edge-on anchoring of the discotic molecules a textural transition from a circular concentric state, unknown from the bulk liquid crystal, to an axial columnar configuration is found accompanied by a change between positive and negative birefringence as a function of decreasing pore size. The peculiar temperature- and confinement-dependent liquid-crystalline texture formation is traced to a complex interplay of different elastic energies, of the anchoring at the pore walls and of the 1-D confining geometry. It allows one to adapt the effective optical anisotropy of the hybrid system in a versatile manner by a proper choice of temperature, pore-size and pore-surface-grafting. 

The onset of collective columnar order and thus of sizeable positive or negative birefringence is shifted compared to the bulk temperature according to the classical Gibbs-Thomson $1/d$-scaling with distinct interfacial energies for the distinct confined textures. This allows one to adjust in a systematic manner by a proper choice of pore size the onset of optical anisotropy.

Instead of disk-like mesogens one may also think about rod-like building blocks in combination with appropriate pore surface-anchoring for adjusting the optical anisotropy of porous solids. However, e.g. for cylindrical nanochannels and homeotropic anchoring such calamatic systems tend to form escape-radial structures \cite{Crawford1996, Kityk2018}, where the director starts at the pore wall with radial order and escapes in the centres in axial directions. This results in competing positive and negative birefringence contributions. Thus the robust columnar stacking principle of discotic liquid crystals, even in extreme spatial confinement, appears as particularly suitable for the establishment of pure radial and axial textures and thus for tailoring optical birefringence in nanoporous media.

We believe that this adjustability is not only of fundamental virtue. The simple functionalisation approach by capillarity-driven spontaneous imbibition of the liquid-crystalline melt in combination with tailorability of anodic aluminium oxide \cite{Lee2014, Chen2015, Sukarno2017, Busch2017} and other self-organized, optically transparent mesoporous media \cite{Gallego2011, Sousa2014, Huber2015, Spengler2018} may have also technological benefits. It allows, for example, integrating a temperature-sensor functionality with optical readout into nanoporous materials. One can also envision that the adjustability of the optical properties by purely tuning collective molecular order along with the mechanical stability of nanoporous media is particularly interesting to induce well-defined lateral phase shift gradients of electromagnetic waves in solid supports or surfaces, e.g. by lateral temperature, magnetic field, or electrical gradients \cite{Yu2011, Pendry2012, Buchnev2015, Busch2017, Nemati2018, ZhangM2018, Shaltout2019}. Such phase gradients are at the core of the emerging field of transformative optics, allowing e.g. negative refraction and the design of adjustable light absorbers or extremely thin metalenses \cite{Yao2014, Lee2018}.    

Thus, similarly as it has been demonstrated for liquid-infused porous structures with regard to confinement-controlled phase selection \cite{Huber2015, Zeng2018}, adaptable surface wetting \cite{Xue2014}, deformation \cite{Gor2015, VanOpdenbosch2016, Gor2017}, topography \cite{Wang2018}, and interactions with biomolecules \cite{Daggumati2015}, our study indicates that liquid-crystal-infused nanoporous solids allow for a quite simple design of materials with adaptable functionality.

From a more general perspective our work is also a fine example, how the combination of functional soft matter with nanostructured porous media allows one to bridge the gap between bottom-up self-assembly of molecular systems with the top-down self-organisation of porosity in monolithic scaffold structures for the design of 3D mechanical robust materials, which is a particular challenge for embedding functional nanocomposites in macroscale devices \cite{Lopez2014, Dreyer2016, Begley2019}.

\section*{Conflict of interest}
There are no conflicts to declare.

\section*{Author contributions}
K.S. and P.H. conceived the experiments. M.L. contributed to design the experiments. K.S. performed the optical and X-ray diffraction experiments and analyzed the data. A.Y. and A.S. supported the sample preparation. M.L., F.B. O.H.S. and P.H. supported the X-ray diffraction experiments. A.W.Z. and M.G.M. performed the Monte Carlo simulations. A.V.K. supported the analysis and interpretation of the optical retardation measurements. K.S., P.H. and M.G.M. wrote the manuscript. All other authors proofread the manuscript.

\section*{Acknowledgements}

Funding by the Deutsche Forschungsgemeinschaft (DFG, German Research Foundation) Projektnummer 192346071, SFB 986 ''Tailor-Made Multi-Scale Materials Systems'', as well as the projects SCHO 470/21-1 and HU 850/5-1 is acknowledged. We thank Deutsche Elektronen-Synchrotron DESY, Hamburg for access to the beamline P08 of the PETRA III synchrotron and the European Synchrotron Radiation Facility (ESRF) for beamtime at ID31. Moreover, we profited from an experiment at the Complex Materials Scattering (CMS) beamline 11-BM of the National Synchrotron Lights Sources NSLS-II, a U.S. Department of Energy (DOE) Office of Science User Facility operated for the DOE Office of Science by Brookhaven National Laboratory under Contract No. DE-SC0012704. The presented results are part of a project that has received funding from the European Union Horizon 2020 research and innovation programme under the Marie Skodowska-Curie grant agreement No 778156. Support from resources for science in years 2018-2022 granted for the realization of  international co-financed project Nr W13/H2020/2018 (Dec. MNiSW 3871/H2020/2018/2) is also acknowledged.
\section*{Dedication}
\noindent Patrick Huber dedicates this work to Prof. Peter S. Pershan (Harvard University), a pioneer in the field of soft condensed matter physics, on his 85th birthday, remembering his mentoring in synchrotron-based X-ray scattering from liquids.





\begin{mcitethebibliography}{96}
\providecommand*{\natexlab}[1]{#1}
\providecommand*{\mciteSetBstSublistMode}[1]{}
\providecommand*{\mciteSetBstMaxWidthForm}[2]{}
\providecommand*{\mciteBstWouldAddEndPuncttrue}
  {\def\EndOfBibitem{\unskip.}}
\providecommand*{\mciteBstWouldAddEndPunctfalse}
  {\let\EndOfBibitem\relax}
\providecommand*{\mciteSetBstMidEndSepPunct}[3]{}
\providecommand*{\mciteSetBstSublistLabelBeginEnd}[3]{}
\providecommand*{\EndOfBibitem}{}
\mciteSetBstSublistMode{f}
\mciteSetBstMaxWidthForm{subitem}
{(\emph{\alph{mcitesubitemcount}})}
\mciteSetBstSublistLabelBeginEnd{\mcitemaxwidthsubitemform\space}
{\relax}{\relax}

\bibitem[Li \emph{et~al.}(2018)Li, Li, Wang, Wang, Liu, Yang, Gong, and
  Gao]{Li2018}
Q.~Li, Z.~Z. Li, X.~Y. Wang, T.~T. Wang, H.~Liu, H.~G. Yang, Y.~Gong and J.~S.
  Gao, \emph{Nanoscale}, 2018, \textbf{10}, 19117--19124\relax
\mciteBstWouldAddEndPuncttrue
\mciteSetBstMidEndSepPunct{\mcitedefaultmidpunct}
{\mcitedefaultendpunct}{\mcitedefaultseppunct}\relax
\EndOfBibitem
\bibitem[Kadic \emph{et~al.}(2019)Kadic, Milton, van Hecke, and
  Wegener]{Kadic2019}
M.~Kadic, G.~W. Milton, M.~van Hecke and M.~Wegener, \emph{Nature Reviews
  Physics}, 2019, \textbf{1}, 198--210\relax
\mciteBstWouldAddEndPuncttrue
\mciteSetBstMidEndSepPunct{\mcitedefaultmidpunct}
{\mcitedefaultendpunct}{\mcitedefaultseppunct}\relax
\EndOfBibitem
\bibitem[Shaltout \emph{et~al.}(2019)Shaltout, Shalaev, and
  Brongersma]{Shaltout2019}
A.~M. Shaltout, V.~M. Shalaev and M.~L. Brongersma, \emph{Science}, 2019,
  \textbf{364}, eaat3100\relax
\mciteBstWouldAddEndPuncttrue
\mciteSetBstMidEndSepPunct{\mcitedefaultmidpunct}
{\mcitedefaultendpunct}{\mcitedefaultseppunct}\relax
\EndOfBibitem
\bibitem[Zheludev and Kivshar(2012)]{Zheludev2012}
N.~I. Zheludev and Y.~S. Kivshar, \emph{Nature Materials}, 2012, \textbf{11},
  917\relax
\mciteBstWouldAddEndPuncttrue
\mciteSetBstMidEndSepPunct{\mcitedefaultmidpunct}
{\mcitedefaultendpunct}{\mcitedefaultseppunct}\relax
\EndOfBibitem
\bibitem[Jalas \emph{et~al.}(2017)Jalas, Shao, Canchi, Okuma, Lang, Petrov,
  Weissm{\"{u}}ller, and Eich]{Jalas2017}
D.~Jalas, L.-H. Shao, R.~Canchi, T.~Okuma, S.~Lang, A.~Petrov,
  J.~Weissm{\"{u}}ller and M.~Eich, \emph{Scientific Reports}, 2017,
  \textbf{7}, 44139\relax
\mciteBstWouldAddEndPuncttrue
\mciteSetBstMidEndSepPunct{\mcitedefaultmidpunct}
{\mcitedefaultendpunct}{\mcitedefaultseppunct}\relax
\EndOfBibitem
\bibitem[Lee \emph{et~al.}(2018)Lee, Yoon, Lee, Yun, Cho, Lee, Kim, Rho, and
  Lee]{Lee2018}
G.~Y. Lee, G.~Yoon, S.~Y. Lee, H.~Yun, J.~Cho, K.~Lee, H.~Kim, J.~Rho and
  B.~Lee, \emph{Nanoscale}, 2018, \textbf{10}, 4237--4245\relax
\mciteBstWouldAddEndPuncttrue
\mciteSetBstMidEndSepPunct{\mcitedefaultmidpunct}
{\mcitedefaultendpunct}{\mcitedefaultseppunct}\relax
\EndOfBibitem
\bibitem[Nemati \emph{et~al.}(2018)Nemati, Wang, Hong, and Teng]{Nemati2018}
A.~Nemati, Q.~Wang, M.~Hong and J.~Teng, \emph{Opto-Electronic Advances}, 2018,
  \textbf{1}, 180009\relax
\mciteBstWouldAddEndPuncttrue
\mciteSetBstMidEndSepPunct{\mcitedefaultmidpunct}
{\mcitedefaultendpunct}{\mcitedefaultseppunct}\relax
\EndOfBibitem
\bibitem[Zhang \emph{et~al.}(2018)Zhang, Pu, Zhang, Guo, He, Ma, Huang, Li, Yu,
  and Luo]{ZhangM2018}
M.~Zhang, M.~Pu, F.~Zhang, Y.~Guo, Q.~He, X.~Ma, Y.~Huang, X.~Li, H.~Yu and
  X.~Luo, \emph{Advanced Science}, 2018, \textbf{5}, 1800835\relax
\mciteBstWouldAddEndPuncttrue
\mciteSetBstMidEndSepPunct{\mcitedefaultmidpunct}
{\mcitedefaultendpunct}{\mcitedefaultseppunct}\relax
\EndOfBibitem
\bibitem[Hu \emph{et~al.}(2019)Hu, Chen, Chen, Guo, Liu, and Dai]{Hu2019}
D.~Hu, K.~Chen, X.~Chen, X.~Guo, M.~Liu and Q.~Dai, \emph{Advanced Materials},
  2019, \textbf{31}, 1807788\relax
\mciteBstWouldAddEndPuncttrue
\mciteSetBstMidEndSepPunct{\mcitedefaultmidpunct}
{\mcitedefaultendpunct}{\mcitedefaultseppunct}\relax
\EndOfBibitem
\bibitem[Matthias \emph{et~al.}(2005)Matthias, Roder, Wehrspohn, Kitzerow,
  Matthias, and Picken]{Matthias2005}
H.~Matthias, T.~Roder, R.~B. Wehrspohn, H.~S. Kitzerow, S.~Matthias and S.~J.
  Picken, \emph{Applied Physics Letters}, 2005, \textbf{87}, 241105\relax
\mciteBstWouldAddEndPuncttrue
\mciteSetBstMidEndSepPunct{\mcitedefaultmidpunct}
{\mcitedefaultendpunct}{\mcitedefaultseppunct}\relax
\EndOfBibitem
\bibitem[Spengler \emph{et~al.}(2018)Spengler, Dong, Michal, Hamad, MacLachlan,
  and Giese]{Spengler2018}
M.~Spengler, R.~Y. Dong, C.~A. Michal, W.~Y. Hamad, M.~J. MacLachlan and
  M.~Giese, \emph{Advanced Functional Materials}, 2018, \textbf{28},
  1800207\relax
\mciteBstWouldAddEndPuncttrue
\mciteSetBstMidEndSepPunct{\mcitedefaultmidpunct}
{\mcitedefaultendpunct}{\mcitedefaultseppunct}\relax
\EndOfBibitem
\bibitem[Buchnev \emph{et~al.}(2015)Buchnev, Podoliak, Kaczmarek, Zheludev, and
  Fedotov]{Buchnev2015}
O.~Buchnev, N.~Podoliak, M.~Kaczmarek, N.~I. Zheludev and V.~A. Fedotov,
  \emph{Advanced Optical Materials}, 2015, \textbf{3}, 674--679\relax
\mciteBstWouldAddEndPuncttrue
\mciteSetBstMidEndSepPunct{\mcitedefaultmidpunct}
{\mcitedefaultendpunct}{\mcitedefaultseppunct}\relax
\EndOfBibitem
\bibitem[Huber(2015)]{Huber2015}
P.~Huber, \emph{J. Phys.: Cond. Matt.}, 2015, \textbf{27}, 103102\relax
\mciteBstWouldAddEndPuncttrue
\mciteSetBstMidEndSepPunct{\mcitedefaultmidpunct}
{\mcitedefaultendpunct}{\mcitedefaultseppunct}\relax
\EndOfBibitem
\bibitem[Ocko \emph{et~al.}(1986)Ocko, Braslau, Pershan, Als-Nielsen, and
  Deutsch]{Ocko1986}
B.~M. Ocko, A.~Braslau, P.~S. Pershan, J.~Als-Nielsen and M.~Deutsch,
  \emph{Phys. Rev. Lett.}, 1986, \textbf{57}, 94\relax
\mciteBstWouldAddEndPuncttrue
\mciteSetBstMidEndSepPunct{\mcitedefaultmidpunct}
{\mcitedefaultendpunct}{\mcitedefaultseppunct}\relax
\EndOfBibitem
\bibitem[Pershan(1988)]{Pershan1988}
P.~Pershan, \emph{Structure of Liquid Crystal Phases}, World Scientific Lecture
  Notes in Physics Vol. 23, World Scientific Publishing, Singapore, 1988\relax
\mciteBstWouldAddEndPuncttrue
\mciteSetBstMidEndSepPunct{\mcitedefaultmidpunct}
{\mcitedefaultendpunct}{\mcitedefaultseppunct}\relax
\EndOfBibitem
\bibitem[Crawford and Zumer(1996)]{Crawford1996}
\emph{Liquid crystals in complex geometries formed by polymer and porous
  networks}, ed. G.~Crawford and S.~Zumer, Taylor and Francis; New York,
  U.S.A., 1996\relax
\mciteBstWouldAddEndPuncttrue
\mciteSetBstMidEndSepPunct{\mcitedefaultmidpunct}
{\mcitedefaultendpunct}{\mcitedefaultseppunct}\relax
\EndOfBibitem
\bibitem[Kutnjak \emph{et~al.}(2003)Kutnjak, Kralj, Lahajnar, and
  Zumer]{Kutnjak2003}
Z.~Kutnjak, S.~Kralj, G.~Lahajnar and S.~Zumer, \emph{Phys. Rev. E}, 2003,
  \textbf{68}, 021705\relax
\mciteBstWouldAddEndPuncttrue
\mciteSetBstMidEndSepPunct{\mcitedefaultmidpunct}
{\mcitedefaultendpunct}{\mcitedefaultseppunct}\relax
\EndOfBibitem
\bibitem[Binder \emph{et~al.}(2008)Binder, Horbach, Vink, and
  De~Virgiliis]{Binder2008}
K.~Binder, J.~Horbach, R.~Vink and A.~De~Virgiliis, \emph{Soft Matter}, 2008,
  \textbf{4}, 1555--1568\relax
\mciteBstWouldAddEndPuncttrue
\mciteSetBstMidEndSepPunct{\mcitedefaultmidpunct}
{\mcitedefaultendpunct}{\mcitedefaultseppunct}\relax
\EndOfBibitem
\bibitem[Mazza \emph{et~al.}(2010)Mazza, Greschek, Valiullin, K{\"{a}}rger, and
  Schoen]{Mazza2010}
M.~G. Mazza, M.~Greschek, R.~Valiullin, J.~K{\"{a}}rger and M.~Schoen,
  \emph{Physical Review Letters}, 2010, \textbf{105}, 227802\relax
\mciteBstWouldAddEndPuncttrue
\mciteSetBstMidEndSepPunct{\mcitedefaultmidpunct}
{\mcitedefaultendpunct}{\mcitedefaultseppunct}\relax
\EndOfBibitem
\bibitem[Chahine \emph{et~al.}(2010)Chahine, Kityk, D\'emarest, Jean, Knorr,
  Huber, Lefort, Zanotti, and Morineau]{Chahine2010}
G.~Chahine, A.~V. Kityk, N.~D\'emarest, F.~Jean, K.~Knorr, P.~Huber, R.~Lefort,
  J.-M. Zanotti and D.~Morineau, \emph{Phys. Rev. E}, 2010, \textbf{82},
  011706\relax
\mciteBstWouldAddEndPuncttrue
\mciteSetBstMidEndSepPunct{\mcitedefaultmidpunct}
{\mcitedefaultendpunct}{\mcitedefaultseppunct}\relax
\EndOfBibitem
\bibitem[Mueter \emph{et~al.}(2010)Mueter, Shin, Deme, Fratzl, Paris, and
  Findenegg]{Mueter2010}
D.~Mueter, T.~Shin, B.~Deme, P.~Fratzl, O.~Paris and G.~H. Findenegg, \emph{The
  Journal of Physical Chemistry Letters}, 2010, \textbf{1}, 1442--1446\relax
\mciteBstWouldAddEndPuncttrue
\mciteSetBstMidEndSepPunct{\mcitedefaultmidpunct}
{\mcitedefaultendpunct}{\mcitedefaultseppunct}\relax
\EndOfBibitem
\bibitem[Araki \emph{et~al.}(2011)Araki, Buscaglia, Bellini, and
  Tanaka]{Araki2011}
T.~Araki, M.~Buscaglia, T.~Bellini and H.~Tanaka, \emph{Nat. Mat.}, 2011,
  \textbf{10}, 303--309\relax
\mciteBstWouldAddEndPuncttrue
\mciteSetBstMidEndSepPunct{\mcitedefaultmidpunct}
{\mcitedefaultendpunct}{\mcitedefaultseppunct}\relax
\EndOfBibitem
\bibitem[Cetinkaya \emph{et~al.}(2013)Cetinkaya, Yildiz, Ozbek, Losada-Perez,
  Leys, and Thoen]{Cetinkaya2013}
M.~C. Cetinkaya, S.~Yildiz, H.~Ozbek, P.~Losada-Perez, J.~Leys and J.~Thoen,
  \emph{Phys. Rev. E}, 2013, \textbf{88}, 042502\relax
\mciteBstWouldAddEndPuncttrue
\mciteSetBstMidEndSepPunct{\mcitedefaultmidpunct}
{\mcitedefaultendpunct}{\mcitedefaultseppunct}\relax
\EndOfBibitem
\bibitem[Calus \emph{et~al.}(2015)Calus, Kityk, Eich, and Huber]{Calus2015}
S.~Calus, A.~V. Kityk, M.~Eich and P.~Huber, \emph{Soft Matter}, 2015,
  \textbf{11}, 3176\relax
\mciteBstWouldAddEndPuncttrue
\mciteSetBstMidEndSepPunct{\mcitedefaultmidpunct}
{\mcitedefaultendpunct}{\mcitedefaultseppunct}\relax
\EndOfBibitem
\bibitem[Schlotthauer \emph{et~al.}(2015)Schlotthauer, Skutnik, Stieger, and
  Schoen]{Schlotthauer2015}
S.~Schlotthauer, R.~A. Skutnik, T.~Stieger and M.~Schoen, \emph{J. Chem.
  Phys.}, 2015, \textbf{142}, 194704\relax
\mciteBstWouldAddEndPuncttrue
\mciteSetBstMidEndSepPunct{\mcitedefaultmidpunct}
{\mcitedefaultendpunct}{\mcitedefaultseppunct}\relax
\EndOfBibitem
\bibitem[Ryu and Yoon(2016)]{Ryu2016}
S.~H. Ryu and D.~K. Yoon, \emph{Liquid Crystals}, 2016, \textbf{43},
  1951--1972\relax
\mciteBstWouldAddEndPuncttrue
\mciteSetBstMidEndSepPunct{\mcitedefaultmidpunct}
{\mcitedefaultendpunct}{\mcitedefaultseppunct}\relax
\EndOfBibitem
\bibitem[Busch \emph{et~al.}(2017)Busch, Kityk, Piecek, Hofmann, Wallacher,
  Ca{\l}us, Kula, Steinhart, Eich, and Huber]{Busch2017}
M.~Busch, A.~V. Kityk, W.~Piecek, T.~Hofmann, D.~Wallacher, S.~Ca{\l}us,
  P.~Kula, M.~Steinhart, M.~Eich and P.~Huber, \emph{Nanoscale}, 2017,
  \textbf{9}, 19086--19099\relax
\mciteBstWouldAddEndPuncttrue
\mciteSetBstMidEndSepPunct{\mcitedefaultmidpunct}
{\mcitedefaultendpunct}{\mcitedefaultseppunct}\relax
\EndOfBibitem
\bibitem[Tran \emph{et~al.}(2017)Tran, Lavrentovich, Durey, Darmon, Haase, Li,
  Lee, Stebe, Kamien, and Lopez-Leon]{Tran2017}
L.~Tran, M.~O. Lavrentovich, G.~Durey, A.~Darmon, M.~F. Haase, N.~Li, D.~Lee,
  K.~J. Stebe, R.~D. Kamien and T.~Lopez-Leon, \emph{Phys. Rev. X}, 2017,
  \textbf{7}, 041029\relax
\mciteBstWouldAddEndPuncttrue
\mciteSetBstMidEndSepPunct{\mcitedefaultmidpunct}
{\mcitedefaultendpunct}{\mcitedefaultseppunct}\relax
\EndOfBibitem
\bibitem[Brumby \emph{et~al.}(2017)Brumby, Wensink, Haslam, and
  Jackson]{Brumby2017}
P.~E. Brumby, H.~H. Wensink, A.~J. Haslam and G.~Jackson, \emph{Langmuir},
  2017, \textbf{33}, 11754--11770\relax
\mciteBstWouldAddEndPuncttrue
\mciteSetBstMidEndSepPunct{\mcitedefaultmidpunct}
{\mcitedefaultendpunct}{\mcitedefaultseppunct}\relax
\EndOfBibitem
\bibitem[Zhang \emph{et~al.}(2019)Zhang, Zeng, Wu, Jin, Liu, and
  Ungar]{Zhang2019}
R.-b. Zhang, X.-b. Zeng, C.~Wu, Q.~Jin, Y.~Liu and G.~Ungar, \emph{Advanced
  Functional Materials}, 2019, \textbf{29}, 1806078\relax
\mciteBstWouldAddEndPuncttrue
\mciteSetBstMidEndSepPunct{\mcitedefaultmidpunct}
{\mcitedefaultendpunct}{\mcitedefaultseppunct}\relax
\EndOfBibitem
\bibitem[Chandrasekhar~R and Ranganath(1990)]{Chandrasekhar1990}
S.~Chandrasekhar~R and G.~S. Ranganath, \emph{Reports On Progress In Physics},
  1990, \textbf{53}, 57--84\relax
\mciteBstWouldAddEndPuncttrue
\mciteSetBstMidEndSepPunct{\mcitedefaultmidpunct}
{\mcitedefaultendpunct}{\mcitedefaultseppunct}\relax
\EndOfBibitem
\bibitem[Oswald and Pieranski(2005)]{Oswald2005}
P.~Oswald and P.~Pieranski, \emph{Smectic and columnar liquid crystals:
  concepts and physical properties illustrated by experiments}, CRC Press, NY,
  2005\relax
\mciteBstWouldAddEndPuncttrue
\mciteSetBstMidEndSepPunct{\mcitedefaultmidpunct}
{\mcitedefaultendpunct}{\mcitedefaultseppunct}\relax
\EndOfBibitem
\bibitem[Feng \emph{et~al.}(2009)Feng, Marcon, Pisula, Hansen, Kirkpatrick,
  Grozema, Andrienko, Kremer, and Muellen]{Feng2009}
X.~Feng, V.~Marcon, W.~Pisula, M.~R. Hansen, J.~Kirkpatrick, F.~Grozema,
  D.~Andrienko, K.~Kremer and K.~Muellen, \emph{Nat. Mater.}, 2009, \textbf{8},
  421--426\relax
\mciteBstWouldAddEndPuncttrue
\mciteSetBstMidEndSepPunct{\mcitedefaultmidpunct}
{\mcitedefaultendpunct}{\mcitedefaultseppunct}\relax
\EndOfBibitem
\bibitem[Sergeyev \emph{et~al.}(2007)Sergeyev, Pisula, and
  Geerts]{Sergeyev2007}
S.~Sergeyev, W.~Pisula and Y.~H. Geerts, \emph{Chem. Soc. Rev.}, 2007,
  \textbf{36}, 1902--1929\relax
\mciteBstWouldAddEndPuncttrue
\mciteSetBstMidEndSepPunct{\mcitedefaultmidpunct}
{\mcitedefaultendpunct}{\mcitedefaultseppunct}\relax
\EndOfBibitem
\bibitem[Bisoyi and Kumar(2010)]{Bisoyi2010}
H.~K. Bisoyi and S.~Kumar, \emph{Chem. Soc. Rev.}, 2010, \textbf{39},
  264--285\relax
\mciteBstWouldAddEndPuncttrue
\mciteSetBstMidEndSepPunct{\mcitedefaultmidpunct}
{\mcitedefaultendpunct}{\mcitedefaultseppunct}\relax
\EndOfBibitem
\bibitem[Kumar(2010)]{Kumar2010}
S.~Kumar, \emph{Chemistry of discotic liquid crystals: from monomers to
  polymers}, CRC Press, NY, 2010\relax
\mciteBstWouldAddEndPuncttrue
\mciteSetBstMidEndSepPunct{\mcitedefaultmidpunct}
{\mcitedefaultendpunct}{\mcitedefaultseppunct}\relax
\EndOfBibitem
\bibitem[Woehrle \emph{et~al.}(2016)Woehrle, Wurzbach, Kirres, Kostidou,
  Kapernaum, Litterscheidt, Haenle, Staffeld, Baro, Giesselmann, and
  Laschat]{Woehrle2015}
T.~Woehrle, I.~Wurzbach, J.~Kirres, A.~Kostidou, N.~Kapernaum,
  J.~Litterscheidt, J.~C. Haenle, P.~Staffeld, A.~Baro, F.~Giesselmann and
  S.~Laschat, \emph{Chem. Rev.}, 2016, \textbf{116}, 1139\relax
\mciteBstWouldAddEndPuncttrue
\mciteSetBstMidEndSepPunct{\mcitedefaultmidpunct}
{\mcitedefaultendpunct}{\mcitedefaultseppunct}\relax
\EndOfBibitem
\bibitem[Bisoyi and Li(2019)]{Bisoyi2019}
H.~K. Bisoyi and Q.~Li, \emph{Progress in Materials Science}, 2019,
  \textbf{104}, 1 -- 52\relax
\mciteBstWouldAddEndPuncttrue
\mciteSetBstMidEndSepPunct{\mcitedefaultmidpunct}
{\mcitedefaultendpunct}{\mcitedefaultseppunct}\relax
\EndOfBibitem
\bibitem[Schmidt-Mende \emph{et~al.}(2001)Schmidt-Mende, Fechtenkotter, Mullen,
  Moons, Friend, and MacKenzie]{Schmidt-Mende2001}
L.~Schmidt-Mende, A.~Fechtenkotter, K.~Mullen, E.~Moons, R.~H. Friend and J.~D.
  MacKenzie, \emph{Science}, 2001, \textbf{293}, 1119--1122\relax
\mciteBstWouldAddEndPuncttrue
\mciteSetBstMidEndSepPunct{\mcitedefaultmidpunct}
{\mcitedefaultendpunct}{\mcitedefaultseppunct}\relax
\EndOfBibitem
\bibitem[Steinhart \emph{et~al.}(2005)Steinhart, Zimmermann, Goring, Schaper,
  Gosele, Weder, and Wendorff]{Steinhart2005}
M.~Steinhart, S.~Zimmermann, P.~Goring, A.~K. Schaper, U.~Gosele, C.~Weder and
  J.~H. Wendorff, \emph{Nano Letters}, 2005, \textbf{5}, 429--434\relax
\mciteBstWouldAddEndPuncttrue
\mciteSetBstMidEndSepPunct{\mcitedefaultmidpunct}
{\mcitedefaultendpunct}{\mcitedefaultseppunct}\relax
\EndOfBibitem
\bibitem[Duran \emph{et~al.}(2012)Duran, Hartmann-Azanza, Steinhart, Gehrig,
  Laquai, Feng, Mullen, Butt, and Floudas]{Duran2012}
H.~Duran, B.~Hartmann-Azanza, M.~Steinhart, D.~Gehrig, F.~Laquai, X.~L. Feng,
  K.~Mullen, H.~J. Butt and G.~Floudas, \emph{ACS Nano}, 2012, \textbf{6},
  9359--9365\relax
\mciteBstWouldAddEndPuncttrue
\mciteSetBstMidEndSepPunct{\mcitedefaultmidpunct}
{\mcitedefaultendpunct}{\mcitedefaultseppunct}\relax
\EndOfBibitem
\bibitem[Kityk \emph{et~al.}(2014)Kityk, Busch, Rau, Calus, Cerclier, Lefort,
  Morineau, Grelet, Krause, Sch{\"o}nhals,\emph{et~al.}]{Kityk2014}
A.~V. Kityk, M.~Busch, D.~Rau, S.~Calus, C.~V. Cerclier, R.~Lefort,
  D.~Morineau, E.~Grelet, C.~Krause, A.~Sch{\"o}nhals \emph{et~al.}, \emph{Soft
  Matter}, 2014, \textbf{10}, 4522--4534\relax
\mciteBstWouldAddEndPuncttrue
\mciteSetBstMidEndSepPunct{\mcitedefaultmidpunct}
{\mcitedefaultendpunct}{\mcitedefaultseppunct}\relax
\EndOfBibitem
\bibitem[Zhang \emph{et~al.}(2017)Zhang, Ungar, Zeng, and Shen]{Zhang2017}
R.~B. Zhang, G.~Ungar, X.~B. Zeng and Z.~H. Shen, \emph{Soft Matter}, 2017,
  \textbf{13}, 4122--4131\relax
\mciteBstWouldAddEndPuncttrue
\mciteSetBstMidEndSepPunct{\mcitedefaultmidpunct}
{\mcitedefaultendpunct}{\mcitedefaultseppunct}\relax
\EndOfBibitem
\bibitem[Cerclier \emph{et~al.}(2012)Cerclier, Ndao, Busselez, Lefort, Grelet,
  Huber, Kityk, Noirez, Schönhals, and Morineau]{Cerclier2012}
C.~V. Cerclier, M.~Ndao, R.~Busselez, R.~Lefort, E.~Grelet, P.~Huber, A.~V.
  Kityk, L.~Noirez, A.~Schönhals and D.~Morineau, \emph{J. Phys. Chem. C},
  2012, \textbf{116}, 18990--18998\relax
\mciteBstWouldAddEndPuncttrue
\mciteSetBstMidEndSepPunct{\mcitedefaultmidpunct}
{\mcitedefaultendpunct}{\mcitedefaultseppunct}\relax
\EndOfBibitem
\bibitem[Zantop(2015)]{Zantop2015}
A.~W. Zantop, \emph{M.Sc. thesis}, Georg-August-Universitaet Goettingen,
  2015\relax
\mciteBstWouldAddEndPuncttrue
\mciteSetBstMidEndSepPunct{\mcitedefaultmidpunct}
{\mcitedefaultendpunct}{\mcitedefaultseppunct}\relax
\EndOfBibitem
\bibitem[Zhang \emph{et~al.}(2015)Zhang, Zeng, Kim, Bushby, Shin, Baker,
  Percec, Leowanawat, and Ungar]{Zhang2015}
R.~Zhang, X.~Zeng, B.~Kim, R.~J. Bushby, K.~Shin, P.~J. Baker, V.~Percec,
  P.~Leowanawat and G.~Ungar, \emph{ACS Nano}, 2015, \textbf{9},
  1759--1766\relax
\mciteBstWouldAddEndPuncttrue
\mciteSetBstMidEndSepPunct{\mcitedefaultmidpunct}
{\mcitedefaultendpunct}{\mcitedefaultseppunct}\relax
\EndOfBibitem
\bibitem[Zhang \emph{et~al.}(2014)Zhang, Zeng, Prehm, Liu, Grimm, Geuss,
  Steinhart, Tschierske, and Ungar]{Zhang2014}
R.~Zhang, X.~Zeng, M.~Prehm, F.~Liu, S.~Grimm, M.~Geuss, M.~Steinhart,
  C.~Tschierske and G.~Ungar, \emph{ACS Nano}, 2014, \textbf{8},
  4500--4509\relax
\mciteBstWouldAddEndPuncttrue
\mciteSetBstMidEndSepPunct{\mcitedefaultmidpunct}
{\mcitedefaultendpunct}{\mcitedefaultseppunct}\relax
\EndOfBibitem
\bibitem[Sentker \emph{et~al.}(2018)Sentker, Zantop, Lippmann, Hofmann, Seeck,
  Kityk, Yildirim, Schönhals, Mazza, and Huber]{Sentker2018}
K.~Sentker, A.~W. Zantop, M.~Lippmann, T.~Hofmann, O.~H. Seeck, A.~V. Kityk,
  A.~Yildirim, A.~Schönhals, M.~G. Mazza and P.~Huber, \emph{Phys. Rev.
  Lett.}, 2018, \textbf{120}, 67801\relax
\mciteBstWouldAddEndPuncttrue
\mciteSetBstMidEndSepPunct{\mcitedefaultmidpunct}
{\mcitedefaultendpunct}{\mcitedefaultseppunct}\relax
\EndOfBibitem
\bibitem[Zhang \emph{et~al.}(2012)Zhang, Liu, Zhang, Tian, and Meng]{Zhang2012}
X.~J. Zhang, X.~X. Liu, X.~H. Zhang, Y.~Tian and Y.~G. Meng, \emph{Liquid
  Crystals}, 2012, \textbf{39}, 1305--1313\relax
\mciteBstWouldAddEndPuncttrue
\mciteSetBstMidEndSepPunct{\mcitedefaultmidpunct}
{\mcitedefaultendpunct}{\mcitedefaultseppunct}\relax
\EndOfBibitem
\bibitem[Gruener and Huber(2011)]{Gruener2011}
S.~Gruener and P.~Huber, \emph{J. Phys. : Cond. Matt.}, 2011, \textbf{23},
  184109\relax
\mciteBstWouldAddEndPuncttrue
\mciteSetBstMidEndSepPunct{\mcitedefaultmidpunct}
{\mcitedefaultendpunct}{\mcitedefaultseppunct}\relax
\EndOfBibitem
\bibitem[Grigoriadis \emph{et~al.}(2011)Grigoriadis, Duran, Steinhart, Kappl,
  and Floudas]{Gri2011}
C.~Grigoriadis, H.~Duran, M.~Steinhart, M.~Kappl and G.~Floudas, \emph{ACS
  Nano}, 2011, \textbf{5}, 9208--9215\relax
\mciteBstWouldAddEndPuncttrue
\mciteSetBstMidEndSepPunct{\mcitedefaultmidpunct}
{\mcitedefaultendpunct}{\mcitedefaultseppunct}\relax
\EndOfBibitem
\bibitem[Yildirim \emph{et~al.}(2019)Yildirim, Sentker, Smales, Pauw, Huber,
  and Schönhals]{Yildirim2019}
A.~Yildirim, K.~Sentker, G.~J. Smales, B.~R. Pauw, P.~Huber and A.~Schönhals,
  \emph{Nanoscale Adv.}, 2019, \textbf{1}, 1104--1116\relax
\mciteBstWouldAddEndPuncttrue
\mciteSetBstMidEndSepPunct{\mcitedefaultmidpunct}
{\mcitedefaultendpunct}{\mcitedefaultseppunct}\relax
\EndOfBibitem
\bibitem[Kityk \emph{et~al.}(2008)Kityk, Wolff, Knorr, Morineau, Lefort, and
  Huber]{Kityk2008}
A.~V. Kityk, M.~Wolff, K.~Knorr, D.~Morineau, R.~Lefort and P.~Huber,
  \emph{Phys. Rev. Lett.}, 2008, \textbf{101}, 187801\relax
\mciteBstWouldAddEndPuncttrue
\mciteSetBstMidEndSepPunct{\mcitedefaultmidpunct}
{\mcitedefaultendpunct}{\mcitedefaultseppunct}\relax
\EndOfBibitem
\bibitem[Kityk \emph{et~al.}(2009)Kityk, Knorr, and Huber]{Kityk2009}
A.~V. Kityk, K.~Knorr and P.~Huber, \emph{Phys. Rev. B}, 2009, \textbf{80},
  035421\relax
\mciteBstWouldAddEndPuncttrue
\mciteSetBstMidEndSepPunct{\mcitedefaultmidpunct}
{\mcitedefaultendpunct}{\mcitedefaultseppunct}\relax
\EndOfBibitem
\bibitem[Note1()]{Note1}
The ESI is available as ancillary file.\relax
\mciteBstWouldAddEndPunctfalse
\mciteSetBstMidEndSepPunct{\mcitedefaultmidpunct}
{}{\mcitedefaultseppunct}\relax
\EndOfBibitem
\bibitem[Ca{\l}us \emph{et~al.}(2014)Ca{\l}us, Kityk, and Huber]{Calus2014}
S.~Ca{\l}us, A.~V. Kityk and P.~Huber, \emph{Microporous and Mesoporous
  Materials}, 2014, \textbf{197}, 26--32\relax
\mciteBstWouldAddEndPuncttrue
\mciteSetBstMidEndSepPunct{\mcitedefaultmidpunct}
{\mcitedefaultendpunct}{\mcitedefaultseppunct}\relax
\EndOfBibitem
\bibitem[Stribeck and N{\"o}chel(2009)]{Stribeck2009}
N.~Stribeck and U.~N{\"o}chel, \emph{Journal of Applied Crystallography}, 2009,
  \textbf{42}, 295--301\relax
\mciteBstWouldAddEndPuncttrue
\mciteSetBstMidEndSepPunct{\mcitedefaultmidpunct}
{\mcitedefaultendpunct}{\mcitedefaultseppunct}\relax
\EndOfBibitem
\bibitem[Note2()]{Note2}
The movies of the ESI are available at TORE, the open access repository for research data 
of Hamburg University of Technology (TUHH), via the link \protect \href
  {https://doi.org/10.15480/336.2515}{https://doi.org/10.15480/336.2515}.\relax
\mciteBstWouldAddEndPunctfalse
\mciteSetBstMidEndSepPunct{\mcitedefaultmidpunct}
{}{\mcitedefaultseppunct}\relax
\EndOfBibitem
\bibitem[Swendsen and Wang(1986)]{swendsenPRL1986}
R.~H. Swendsen and J.-S. Wang, \emph{Phys. Rev. Lett.}, 1986, \textbf{57},
  2607\relax
\mciteBstWouldAddEndPuncttrue
\mciteSetBstMidEndSepPunct{\mcitedefaultmidpunct}
{\mcitedefaultendpunct}{\mcitedefaultseppunct}\relax
\EndOfBibitem
\bibitem[Bates and Luckhurst(1996)]{BatesJCP1996}
M.~A. Bates and G.~R. Luckhurst, \emph{J. Chem. Phys.}, 1996, \textbf{104},
  6696--6709\relax
\mciteBstWouldAddEndPuncttrue
\mciteSetBstMidEndSepPunct{\mcitedefaultmidpunct}
{\mcitedefaultendpunct}{\mcitedefaultseppunct}\relax
\EndOfBibitem
\bibitem[Yan and de~Pablo(1999)]{Yan1999}
Q.~Yan and J.~J. de~Pablo, \emph{J. Chem. Phys.}, 1999, \textbf{111},
  9509--9516\relax
\mciteBstWouldAddEndPuncttrue
\mciteSetBstMidEndSepPunct{\mcitedefaultmidpunct}
{\mcitedefaultendpunct}{\mcitedefaultseppunct}\relax
\EndOfBibitem
\bibitem[Caprion \emph{et~al.}(2003)Caprion, Bellier-Castella, and
  Ryckaert]{Caprion2003}
D.~Caprion, L.~Bellier-Castella and J.-P. Ryckaert, \emph{Phys. Rev. E}, 2003,
  \textbf{67}, 041703\relax
\mciteBstWouldAddEndPuncttrue
\mciteSetBstMidEndSepPunct{\mcitedefaultmidpunct}
{\mcitedefaultendpunct}{\mcitedefaultseppunct}\relax
\EndOfBibitem
\bibitem[Earl and Deem(2005)]{Earl2005}
D.~J. Earl and M.~W. Deem, \emph{Phys. Chem. Chem. Phys.}, 2005, \textbf{7},
  3910--3916\relax
\mciteBstWouldAddEndPuncttrue
\mciteSetBstMidEndSepPunct{\mcitedefaultmidpunct}
{\mcitedefaultendpunct}{\mcitedefaultseppunct}\relax
\EndOfBibitem
\bibitem[Lechner and Dellago(2008)]{Lechner2008}
W.~Lechner and C.~Dellago, \emph{J. Chem. Phys.}, 2008, \textbf{129},
  114707\relax
\mciteBstWouldAddEndPuncttrue
\mciteSetBstMidEndSepPunct{\mcitedefaultmidpunct}
{\mcitedefaultendpunct}{\mcitedefaultseppunct}\relax
\EndOfBibitem
\bibitem[Caprion(2009)]{caprion2009discotic}
D.~Caprion, \emph{Eur. Phys. J. E}, 2009, \textbf{28}, 305--313\relax
\mciteBstWouldAddEndPuncttrue
\mciteSetBstMidEndSepPunct{\mcitedefaultmidpunct}
{\mcitedefaultendpunct}{\mcitedefaultseppunct}\relax
\EndOfBibitem
\bibitem[Note3()]{Note3}
Note that we observe a second set of Bragg reflections at $\chi $s typical of
  the $(100)_{\perp }$ domain for the sample with 38\protect \tmspace
  +\thinmuskip {.1667em}nm pore size. It appears at temperatures much higher
  than for the confined liquid crystal, close to the bulk transition, and
  exhibits very sharp Bragg reflections indicating large coherence lengths.
  Therefore, we attribute these reflections to a textured bulk film growing
  with one (100) plane set parallel to the AAO membrane surface, rather than
  tracing them to a $(100)_{\perp }$ domain in pore space. However, we cannot
  entirely exclude the latter possibility.\relax
\mciteBstWouldAddEndPunctfalse
\mciteSetBstMidEndSepPunct{\mcitedefaultmidpunct}
{}{\mcitedefaultseppunct}\relax
\EndOfBibitem
\bibitem[Henschel \emph{et~al.}(2007)Henschel, Hofmann, Huber, and
  Knorr]{Henschel2007}
A.~Henschel, T.~Hofmann, P.~Huber and K.~Knorr, \emph{Phys. Rev. E}, 2007,
  \textbf{75}, 021607\relax
\mciteBstWouldAddEndPuncttrue
\mciteSetBstMidEndSepPunct{\mcitedefaultmidpunct}
{\mcitedefaultendpunct}{\mcitedefaultseppunct}\relax
\EndOfBibitem
\bibitem[Henschel \emph{et~al.}(2009)Henschel, Hofmann, Kumar, Knorr, and
  Huber]{Henschel2009}
A.~Henschel, T.~Hofmann, P.~Kumar, K.~Knorr and P.~Huber, \emph{Phys. Rev. E},
  2009, \textbf{79}, 032601--1--4\relax
\mciteBstWouldAddEndPuncttrue
\mciteSetBstMidEndSepPunct{\mcitedefaultmidpunct}
{\mcitedefaultendpunct}{\mcitedefaultseppunct}\relax
\EndOfBibitem
\bibitem[Knorr \emph{et~al.}(2010)Knorr, Huber, and Wallacher]{Knorr2009}
K.~Knorr, P.~Huber and D.~Wallacher, \emph{Z. Phys. Chem.}, 2010, \textbf{222},
  257--285\relax
\mciteBstWouldAddEndPuncttrue
\mciteSetBstMidEndSepPunct{\mcitedefaultmidpunct}
{\mcitedefaultendpunct}{\mcitedefaultseppunct}\relax
\EndOfBibitem
\bibitem[Steinr{\"{u}}ck \emph{et~al.}(2014)Steinr{\"{u}}ck, Magerl, Deutsch,
  and Ocko]{Steinruck2014}
H.~G. Steinr{\"{u}}ck, A.~Magerl, M.~Deutsch and B.~M. Ocko, \emph{Phys. Rev.
  Lett.}, 2014, \textbf{113}, 156101\relax
\mciteBstWouldAddEndPuncttrue
\mciteSetBstMidEndSepPunct{\mcitedefaultmidpunct}
{\mcitedefaultendpunct}{\mcitedefaultseppunct}\relax
\EndOfBibitem
\bibitem[Khassanov \emph{et~al.}(2015)Khassanov, Steinr{\"{u}}ck, Schmaltz,
  Magerl, and Halik]{Khassanov2015}
A.~Khassanov, H.~G. Steinr{\"{u}}ck, T.~Schmaltz, A.~Magerl and M.~Halik,
  \emph{Accounts of Chemical Research}, 2015, \textbf{48}, 1901--1908\relax
\mciteBstWouldAddEndPuncttrue
\mciteSetBstMidEndSepPunct{\mcitedefaultmidpunct}
{\mcitedefaultendpunct}{\mcitedefaultseppunct}\relax
\EndOfBibitem
\bibitem[Klauk \emph{et~al.}(2007)Klauk, Zschieschang, Pflaum, and
  Halik]{Klauk2007}
H.~Klauk, U.~Zschieschang, J.~Pflaum and M.~Halik, \emph{Nature}, 2007,
  \textbf{445}, 745--748\relax
\mciteBstWouldAddEndPuncttrue
\mciteSetBstMidEndSepPunct{\mcitedefaultmidpunct}
{\mcitedefaultendpunct}{\mcitedefaultseppunct}\relax
\EndOfBibitem
\bibitem[Webber \emph{et~al.}(2007)Webber, Dore, Strange, Anderson, and
  Tohidi]{Webber2007}
J.~B.~W. Webber, J.~C. Dore, J.~H. Strange, R.~Anderson and B.~Tohidi,
  \emph{Journal of Physics: Condensed Matter}, 2007, \textbf{19}, 415117\relax
\mciteBstWouldAddEndPuncttrue
\mciteSetBstMidEndSepPunct{\mcitedefaultmidpunct}
{\mcitedefaultendpunct}{\mcitedefaultseppunct}\relax
\EndOfBibitem
\bibitem[Huber and Knorr(1999)]{Huber1999}
P.~Huber and K.~Knorr, \emph{Phys. Rev. B}, 1999, \textbf{60}, 12657\relax
\mciteBstWouldAddEndPuncttrue
\mciteSetBstMidEndSepPunct{\mcitedefaultmidpunct}
{\mcitedefaultendpunct}{\mcitedefaultseppunct}\relax
\EndOfBibitem
\bibitem[Schaefer \emph{et~al.}(2008)Schaefer, Hofmann, Wallacher, Huber, and
  Knorr]{Schaefer2008}
C.~Schaefer, T.~Hofmann, D.~Wallacher, P.~Huber and K.~Knorr, \emph{Phys. Rev.
  Lett.}, 2008, \textbf{100}, 175701\relax
\mciteBstWouldAddEndPuncttrue
\mciteSetBstMidEndSepPunct{\mcitedefaultmidpunct}
{\mcitedefaultendpunct}{\mcitedefaultseppunct}\relax
\EndOfBibitem
\bibitem[Saito and Miyagi(1989)]{Saito1989}
M.~Saito and M.~Miyagi, \emph{J. Opt. Soc. Am. A}, 1989, \textbf{6},
  1895--1900\relax
\mciteBstWouldAddEndPuncttrue
\mciteSetBstMidEndSepPunct{\mcitedefaultmidpunct}
{\mcitedefaultendpunct}{\mcitedefaultseppunct}\relax
\EndOfBibitem
\bibitem[Gong \emph{et~al.}(2011)Gong, Stolz, Myeong, Dogheche, Gokarna, Ryu,
  Decoster, and Cho]{Gong2011}
S.-H. Gong, A.~Stolz, G.-H. Myeong, E.~Dogheche, A.~Gokarna, S.-W. Ryu,
  D.~Decoster and Y.-H. Cho, \emph{Optics letters}, 2011, \textbf{36},
  4272--4\relax
\mciteBstWouldAddEndPuncttrue
\mciteSetBstMidEndSepPunct{\mcitedefaultmidpunct}
{\mcitedefaultendpunct}{\mcitedefaultseppunct}\relax
\EndOfBibitem
\bibitem[Kityk \emph{et~al.}(2018)Kityk, Huber, Sentker, Andrushchak, Kula,
  Piecek, Wielgosz, Kityk, Goering, and Lelonek]{Kityk2018}
A.~V. Kityk, P.~Huber, K.~Sentker, A.~Andrushchak, P.~Kula, W.~Piecek,
  R.~Wielgosz, O.~Kityk, P.~Goering and M.~Lelonek, \textit{arXiv preprint
  arXiv}:1806.05630, 2018\relax
\mciteBstWouldAddEndPuncttrue
\mciteSetBstMidEndSepPunct{\mcitedefaultmidpunct}
{\mcitedefaultendpunct}{\mcitedefaultseppunct}\relax
\EndOfBibitem
\bibitem[Lee and Park(2014)]{Lee2014}
W.~Lee and S.-J. Park, \emph{Chemical Reviews}, 2014, \textbf{114},
  7487--7556\relax
\mciteBstWouldAddEndPuncttrue
\mciteSetBstMidEndSepPunct{\mcitedefaultmidpunct}
{\mcitedefaultendpunct}{\mcitedefaultseppunct}\relax
\EndOfBibitem
\bibitem[Chen \emph{et~al.}(2015)Chen, Santos, Wang, Kumeria, Wang, Li, and
  Losic]{Chen2015}
Y.~Chen, A.~Santos, Y.~Wang, T.~Kumeria, C.~Wang, J.~Li and D.~Losic,
  \emph{Nanoscale}, 2015, \textbf{7}, 7770--7779\relax
\mciteBstWouldAddEndPuncttrue
\mciteSetBstMidEndSepPunct{\mcitedefaultmidpunct}
{\mcitedefaultendpunct}{\mcitedefaultseppunct}\relax
\EndOfBibitem
\bibitem[Sukarno and Santos(2017)]{Sukarno2017}
C.~S.~L. Sukarno and A.~Santos, \emph{Nanoscale}, 2017, \textbf{9},
  7541--7550\relax
\mciteBstWouldAddEndPuncttrue
\mciteSetBstMidEndSepPunct{\mcitedefaultmidpunct}
{\mcitedefaultendpunct}{\mcitedefaultseppunct}\relax
\EndOfBibitem
\bibitem[Gallego-Gomez \emph{et~al.}({2011})Gallego-Gomez, Blanco,
  Canalejas-Tejero, and Lopez]{Gallego2011}
F.~Gallego-Gomez, A.~Blanco, V.~Canalejas-Tejero and C.~Lopez, \emph{{Small}},
  {2011}, \textbf{{7}}, {1838--1845}\relax
\mciteBstWouldAddEndPuncttrue
\mciteSetBstMidEndSepPunct{\mcitedefaultmidpunct}
{\mcitedefaultendpunct}{\mcitedefaultseppunct}\relax
\EndOfBibitem
\bibitem[Sousa \emph{et~al.}(2014)Sousa, Leitao, Proenca, Ventura, Pereira, and
  Araujo]{Sousa2014}
C.~T. Sousa, D.~C. Leitao, M.~P. Proenca, J.~Ventura, A.~M. Pereira and J.~P.
  Araujo, \emph{Applied Physics Reviews}, 2014, \textbf{1},
  031102--1--031102--22\relax
\mciteBstWouldAddEndPuncttrue
\mciteSetBstMidEndSepPunct{\mcitedefaultmidpunct}
{\mcitedefaultendpunct}{\mcitedefaultseppunct}\relax
\EndOfBibitem
\bibitem[Yu \emph{et~al.}(2011)Yu, Genevet, Kats, Aieta, Tetienne, Capasso, and
  Gaburro]{Yu2011}
N.~Yu, P.~Genevet, M.~A. Kats, F.~Aieta, J.-P. Tetienne, F.~Capasso and
  Z.~Gaburro, \emph{Science}, 2011, \textbf{334}, 333--337\relax
\mciteBstWouldAddEndPuncttrue
\mciteSetBstMidEndSepPunct{\mcitedefaultmidpunct}
{\mcitedefaultendpunct}{\mcitedefaultseppunct}\relax
\EndOfBibitem
\bibitem[Pendry \emph{et~al.}(2012)Pendry, Aubry, Smith, and Maier]{Pendry2012}
J.~B. Pendry, A.~Aubry, D.~R. Smith and S.~A. Maier, \emph{Science}, 2012,
  \textbf{337}, 549--552\relax
\mciteBstWouldAddEndPuncttrue
\mciteSetBstMidEndSepPunct{\mcitedefaultmidpunct}
{\mcitedefaultendpunct}{\mcitedefaultseppunct}\relax
\EndOfBibitem
\bibitem[Yao \emph{et~al.}(2014)Yao, Shankar, Kats, Song, Kong, Loncar, and
  Capasso]{Yao2014}
Y.~Yao, R.~Shankar, M.~A. Kats, Y.~Song, J.~Kong, M.~Loncar and F.~Capasso,
  \emph{Nano Letters}, 2014, \textbf{14}, 6526--6532\relax
\mciteBstWouldAddEndPuncttrue
\mciteSetBstMidEndSepPunct{\mcitedefaultmidpunct}
{\mcitedefaultendpunct}{\mcitedefaultseppunct}\relax
\EndOfBibitem
\bibitem[Zeng \emph{et~al.}(2018)Zeng, Kim, Anduix-Canto, Frontera, Laundy,
  Kapur, Christenson, and Meldrum]{Zeng2018}
M.~Zeng, Y.-Y. Kim, C.~Anduix-Canto, C.~Frontera, D.~Laundy, N.~Kapur, H.~K.
  Christenson and F.~C. Meldrum, \emph{Proceedings of the National Academy of
  Sciences}, 2018, \textbf{115}, 7670--7675\relax
\mciteBstWouldAddEndPuncttrue
\mciteSetBstMidEndSepPunct{\mcitedefaultmidpunct}
{\mcitedefaultendpunct}{\mcitedefaultseppunct}\relax
\EndOfBibitem
\bibitem[Xue \emph{et~al.}(2014)Xue, Markmann, Duan, Wei{\ss}m{\"u}ller, and
  Huber]{Xue2014}
Y.~Xue, J.~Markmann, H.~Duan, J.~Wei{\ss}m{\"u}ller and P.~Huber, \emph{Nature
  Communications}, 2014, \textbf{5}, 4237\relax
\mciteBstWouldAddEndPuncttrue
\mciteSetBstMidEndSepPunct{\mcitedefaultmidpunct}
{\mcitedefaultendpunct}{\mcitedefaultseppunct}\relax
\EndOfBibitem
\bibitem[Gor \emph{et~al.}(2015)Gor, Bertinetti, Bernstein, Hofmann, Fratzl,
  and Huber]{Gor2015}
G.~Y. Gor, L.~Bertinetti, N.~Bernstein, T.~Hofmann, P.~Fratzl and P.~Huber,
  \emph{Applied Physics Letters}, 2015, \textbf{106}, 261901\relax
\mciteBstWouldAddEndPuncttrue
\mciteSetBstMidEndSepPunct{\mcitedefaultmidpunct}
{\mcitedefaultendpunct}{\mcitedefaultseppunct}\relax
\EndOfBibitem
\bibitem[Van~Opdenbosch \emph{et~al.}(2016)Van~Opdenbosch, Fritz-Popovski,
  Wagermaier, Paris, and Zollfrank]{VanOpdenbosch2016}
D.~Van~Opdenbosch, G.~Fritz-Popovski, W.~Wagermaier, O.~Paris and C.~Zollfrank,
  \emph{Advanced Materials}, 2016, \textbf{28}, 5235--5240\relax
\mciteBstWouldAddEndPuncttrue
\mciteSetBstMidEndSepPunct{\mcitedefaultmidpunct}
{\mcitedefaultendpunct}{\mcitedefaultseppunct}\relax
\EndOfBibitem
\bibitem[Gor \emph{et~al.}(2017)Gor, Huber, and Bernstein]{Gor2017}
G.~Y. Gor, P.~Huber and N.~Bernstein, \emph{Appl. Phys. Rev.}, 2017,
  \textbf{4}, 011303\relax
\mciteBstWouldAddEndPuncttrue
\mciteSetBstMidEndSepPunct{\mcitedefaultmidpunct}
{\mcitedefaultendpunct}{\mcitedefaultseppunct}\relax
\EndOfBibitem
\bibitem[Wang \emph{et~al.}(2018)Wang, Timonen, Carlson, Drotlef, Zhang, Kolle,
  Grinthal, Wong, Hatton, Kang, Kennedy, Chi, Blough, Sitti, Mahadevan, and
  Aizenberg]{Wang2018}
W.~Wang, J.~V.~I. Timonen, A.~Carlson, D.-M. Drotlef, C.~T. Zhang, S.~Kolle,
  A.~Grinthal, T.-S. Wong, B.~Hatton, S.~H. Kang, S.~Kennedy, J.~Chi, R.~T.
  Blough, M.~Sitti, L.~Mahadevan and J.~Aizenberg, \emph{Nature}, 2018,
  \textbf{559}, 77--82\relax
\mciteBstWouldAddEndPuncttrue
\mciteSetBstMidEndSepPunct{\mcitedefaultmidpunct}
{\mcitedefaultendpunct}{\mcitedefaultseppunct}\relax
\EndOfBibitem
\bibitem[Daggumati \emph{et~al.}(2015)Daggumati, Matharu, Wang, and
  Seker]{Daggumati2015}
P.~Daggumati, Z.~Matharu, L.~Wang and E.~Seker, \emph{Analytical Chemistry},
  2015, \textbf{87}, 8618--8622\relax
\mciteBstWouldAddEndPuncttrue
\mciteSetBstMidEndSepPunct{\mcitedefaultmidpunct}
{\mcitedefaultendpunct}{\mcitedefaultseppunct}\relax
\EndOfBibitem
\bibitem[Lopez-Andarias \emph{et~al.}({2014})Lopez-Andarias, Luis~Lopez,
  Atienza, Brunetti, Romero-Nieto, Guldi, and Martin]{Lopez2014}
J.~Lopez-Andarias, J.~Luis~Lopez, C.~Atienza, F.~G. Brunetti, C.~Romero-Nieto,
  D.~M. Guldi and N.~Martin, \emph{{Nature Communications}}, {2014},
  \textbf{{5}}, {3763}\relax
\mciteBstWouldAddEndPuncttrue
\mciteSetBstMidEndSepPunct{\mcitedefaultmidpunct}
{\mcitedefaultendpunct}{\mcitedefaultseppunct}\relax
\EndOfBibitem
\bibitem[Dreyer \emph{et~al.}(2016)Dreyer, Feld, Kornowski, Yilmaz, Noei,
  Meyer, Krekeler, Jiao, Stierle, Abetz, Weller, and Schneider]{Dreyer2016}
A.~Dreyer, A.~Feld, A.~Kornowski, E.~D. Yilmaz, H.~Noei, A.~Meyer, T.~Krekeler,
  C.~Jiao, A.~Stierle, V.~Abetz, H.~Weller and G.~A. Schneider, \emph{Nature
  Materials}, 2016, \textbf{15}, 522\relax
\mciteBstWouldAddEndPuncttrue
\mciteSetBstMidEndSepPunct{\mcitedefaultmidpunct}
{\mcitedefaultendpunct}{\mcitedefaultseppunct}\relax
\EndOfBibitem
\bibitem[Begley \emph{et~al.}(2019)Begley, Gianola, and Ray]{Begley2019}
M.~R. Begley, D.~S. Gianola and T.~R. Ray, \emph{Science}, 2019, \textbf{364},
  eaav4299\relax
\mciteBstWouldAddEndPuncttrue
\mciteSetBstMidEndSepPunct{\mcitedefaultmidpunct}
{\mcitedefaultendpunct}{\mcitedefaultseppunct}\relax
\EndOfBibitem
\end{mcitethebibliography}
\providecommand*{\mcitethebibliography}{\thebibliography}
\csname @ifundefined\endcsname{endmcitethebibliography}
{\let\endmcitethebibliography\endthebibliography}{}

\end{document}